\documentclass[sigconf,natbib,9pt]{acmart}

\usepackage[subtle, wordspacing=normal, tracking=normal, bibnotes, charwidths=normal, leading, indent, lists, paragraphs, mathspacing=normal, bibliography=normal]{savetrees}
\usepackage{booktabs} 
\usepackage{graphicx}
\usepackage{subcaption}
\usepackage{url}
\usepackage{microtype}
\usepackage{tabularx,amsmath}
\usepackage{amssymb}
\usepackage{multirow}
\usepackage{color}
\usepackage[normalem]{ulem}
\usepackage[english]{babel}
\usepackage{colortbl}
\usepackage[noend]{algorithmic}
\usepackage{algorithm}
\usepackage{paralist}
\usepackage{enumerate}
\usepackage{enumitem}
\usepackage{acronym}
\usepackage[skip=0pt]{caption}
\usepackage{accents}
\usepackage{setspace}
\usepackage{balance}
\usepackage[normalem]{ulem}
\usepackage{bbm}

\setcitestyle{square,comma,numbers,sort&compress,sectionbib}

\providecommand{\argmax}{\operatornamewithlimits{arg\,max}}

\newbox\itembox
\def\itemlistlabel#1{#1\hfill}
\def\itemlist#1{\setbox\itembox=\hbox{#1}%
                \list{}{\labelwidth\wd\itembox
                             \leftmargin\labelwidth
                             \advance\leftmargin by\itemindent
                             \advance\leftmargin by\labelsep
                             \let\makelabel\itemlistlabel}}

\newcommand{\enkelop}{$^{\vartriangle}$}
\newcommand{\dubbelop}{$^{\blacktriangle}$}
\newcommand{\enkelneer}{$^{\triangledown}$}
\newcommand{\dubbelneer}{$^{\blacktriangledown}$}

\acrodef{IR}{Information Retrieval}
\acrodef{RL}{Reinforcement Learning}
\acrodef{LTR}{Learning to Rank}
\acrodef{OLTR}{Online Learning to Rank}
\acrodef{MDP}{Markov Decision Processes}
\acrodef{DQN}{Deep Q-Network}
\acrodef{DCG}{Discounted Cumulative Gain}
\acrodef{RNN}{Recurrent Neural Network}
\acrodef{GRU}{Gated Recurrent Unit Network}
\acrodef{LSTM}{Long Short-Term Memory Network}
\acrodef{DRM}{Double-Rank Model}
\acrodef{ReLU}{Rectified Linear Unit}
\acrodef{P-NDCG}{Permuted Normalized DCG}

\newcommand{\OurModel}{\ac{DRM}}

\title{Ranking for Relevance and Display Preferences\\ in Complex Presentation Layouts}

\author{Harrie Oosterhuis}
\orcid{}
\affiliation{%
\institution{University of Amsterdam}
\city{Amsterdam}
\country{The Netherlands}
}
\email{oosterhuis@uva.nl}

\author{Maarten de Rijke}
\orcid{0000-0002-1086-0202}
\affiliation{%
\institution{University of Amsterdam}
\city{Amsterdam}
\country{The Netherlands}
}
\email{derijke@uva.nl}

\begin{document}

\begin{abstract}
\acl{LTR} has traditionally considered settings where given the relevance information of objects, the desired order in which to rank the objects is clear.
However, with today's large variety of users and layouts this is not always the case.
In this paper, we consider so-called complex ranking settings where it is not clear what should be displayed, that is, what the relevant items are, and how they should be displayed, that is, where the most relevant items should be placed.
These ranking settings are complex as they involve both traditional ranking and inferring the best display order.
Existing learning to rank methods cannot handle such complex ranking settings as they assume that the display order is known beforehand. To address this gap we introduce a novel Deep \acl{RL} method that is capable of learning complex rankings, both the layout and the best ranking given the layout, from weak reward signals.
Our proposed method does so by selecting documents and positions sequentially, hence it ranks both the documents and positions, which is why we call it the \OurModel{}.
Our experiments show that \OurModel{} outperforms all existing methods in complex ranking settings, thus it leads to substantial ranking improvements in cases where the display order is not known a priori.
\end{abstract}

%\begin{CCSXML}
%<ccs2012>
%<concept>
%<concept_id>10002951.10003317.10003338.10003343</concept_id>
%<concept_desc>Information systems~Learning to rank</concept_desc>
%<concept_significance>500</concept_significance>
%</concept>
%<concept>
%<concept_id>10002951.10003317.10003338</concept_id>
%<concept_desc>Information systems~Retrieval models and ranking</concept_desc>
%<concept_significance>300</concept_significance>
%</concept>
%</ccs2012>
%\end{CCSXML}
%
%\ccsdesc[500]{Information systems~Learning to rank}
%\ccsdesc[300]{Information systems~Retrieval models and ranking}


\copyrightyear{2018} 
\acmYear{2018} 
\setcopyright{acmlicensed}
\acmConference[SIGIR '18]{The 41st International ACM SIGIR Conference on Research \& Development in Information Retrieval}{July 8--12, 2018}{Ann Arbor, MI, USA}
\acmBooktitle{SIGIR '18: The 41st International ACM SIGIR Conference on Research \& Development in Information Retrieval, July 8--12, 2018, Ann Arbor, MI, USA}
\acmPrice{15.00}
\acmDOI{10.1145/3209978.3209992}
\acmISBN{978-1-4503-5657-2/18/07}

\maketitle

% !TEX root = sigir2018-complex-rankings.tex

\section{Introduction}
\label{sec:intro}
\acf{LTR} has played a vital role in the field of \acf{IR}. 
It allows search engines to provide users with documents relevant to their search task by carefully combining a large number of ranking signals~\citep{liu2009learning}. 
Similarly, it is an important part of many recommender systems~\citep{karatzoglou-learning-2013}, enables smart advertisement placement~\citep{tagami-ctr-2013}, and is used for effective product search~\citep{karmakersantu-application-2017}. 
Over time, \ac{LTR} has spread to many cases beyond the traditional web search setting, and so have the ways in which users interact with rankings. 
Besides the well-known \emph{ten blue links} result presentation format, a myriad of different layouts are now prevalent on the web. 
For comparison, consider the traditional layout in Fig.~\ref{fig:tenbluelinks}; eye-tracking studies have demonstrated that users look at the top left of such a layout first, before making their way down the list, the so-called ``F-shape'' or ``golden triangle''~\citep{hotchkiss-eye-2005}.
Because of this top-down bias, a more relevant document should be placed higher; correspondingly, \ac{LTR} methods have always relied on this assumption.

In contrast, Fig.~\ref{fig:complexexamples} displays three examples of layouts where the top-down traversal assumption does not necessarily hold. 
First, Fig.~\ref{fig:productsearch} displays a common layout for search in online stores; here, products are presented in a grid so that their name, thumbnail, price and rating can be displayed compactly and navigational  panels are displayed on the left-hand side. Depending on their information need, users are often drawn to the navigational panel, especially when confronted with a large number of results; users are also often drawn to the thumbnail photos that represent products, which they do not seem to scan in a left-to-right, top-to-bottom fashion~\citep{redwood-eye-2009,campion-amazon-2013}.
A typical video recommendation result page is shown in Fig.~\ref{fig:videosearch}. 
Often, video thumbnails are displayed in horizontal strips where each strip shows videos of the same category.
This allows a large number of videos to be shown while still having a structured presentation. 
Users of such result displays tend to have ``T-shaped'' fixation patterns~\citep{xie-investigating-2017,xie-why-2018}, where users view the top-center area first, the top-left area second, and the center-center area third~\citep{lu-eye-tracking-2014}.
Lastly, Fig.~\ref{fig:advertisements} shows common advertisement placements on web sites; ad placement areas vary greatly~\citep{michailidou-towards-2014}, but common ones are top, left rail and right rail, with a mix of shapes.
While little is known about interaction patterns with ads on web sites, studies on ads placed on web search engine result pages show that top and right rail ads receive a higher fraction of visual attention~\citep{buscher-good-2010}, as users have a bias against sponsored links \citep{jansen-factors-2007}.
It is very hard to anticipate where an advertisement would be the most effective as it depends on the overall content of the website.

\begin{figure*}[tb]
\centering
\begin{subfigure}{0.4\textwidth}
\includegraphics[width=\textwidth]{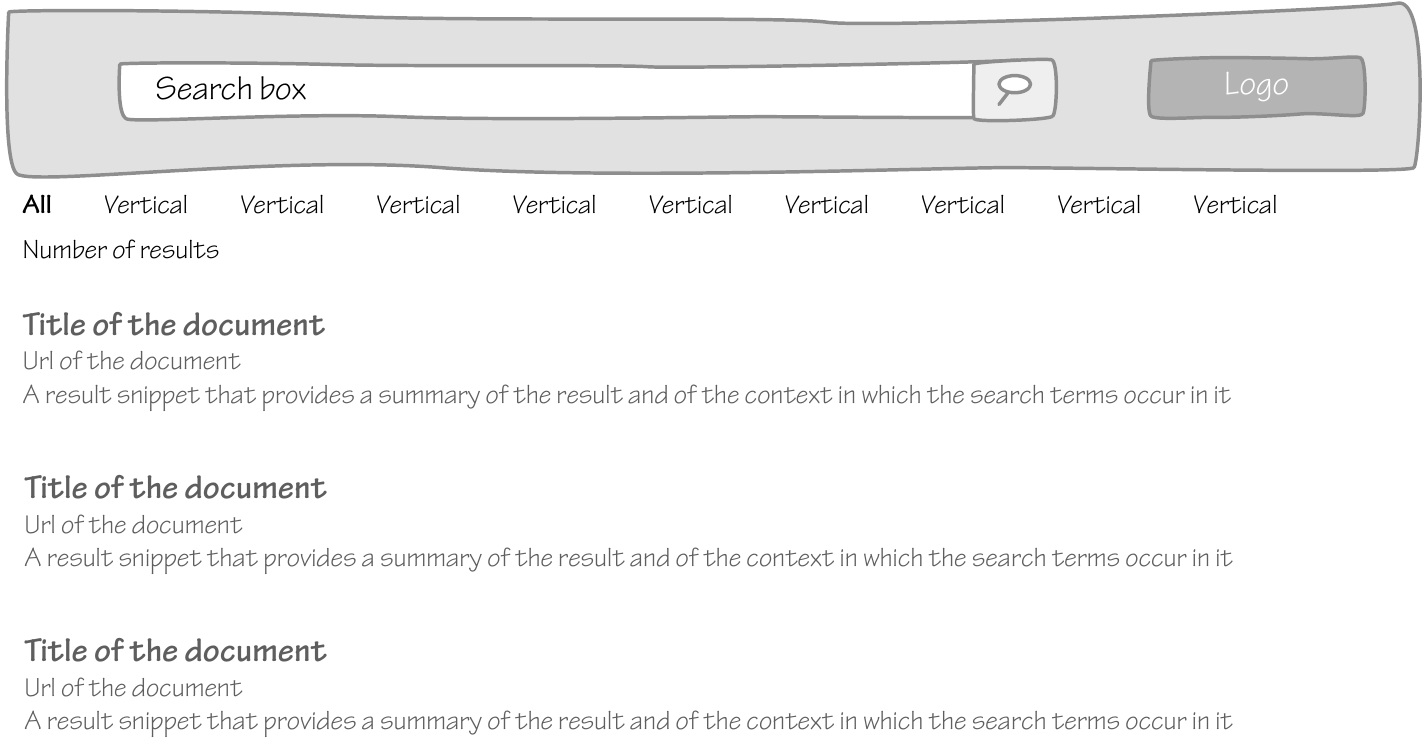}
\caption{A traditional \emph{ten blue links} layout.}
\label{fig:tenbluelinks}
\end{subfigure}
\hspace*{1.5cm}
\begin{subfigure}{0.4\textwidth}
\includegraphics[width=\textwidth]{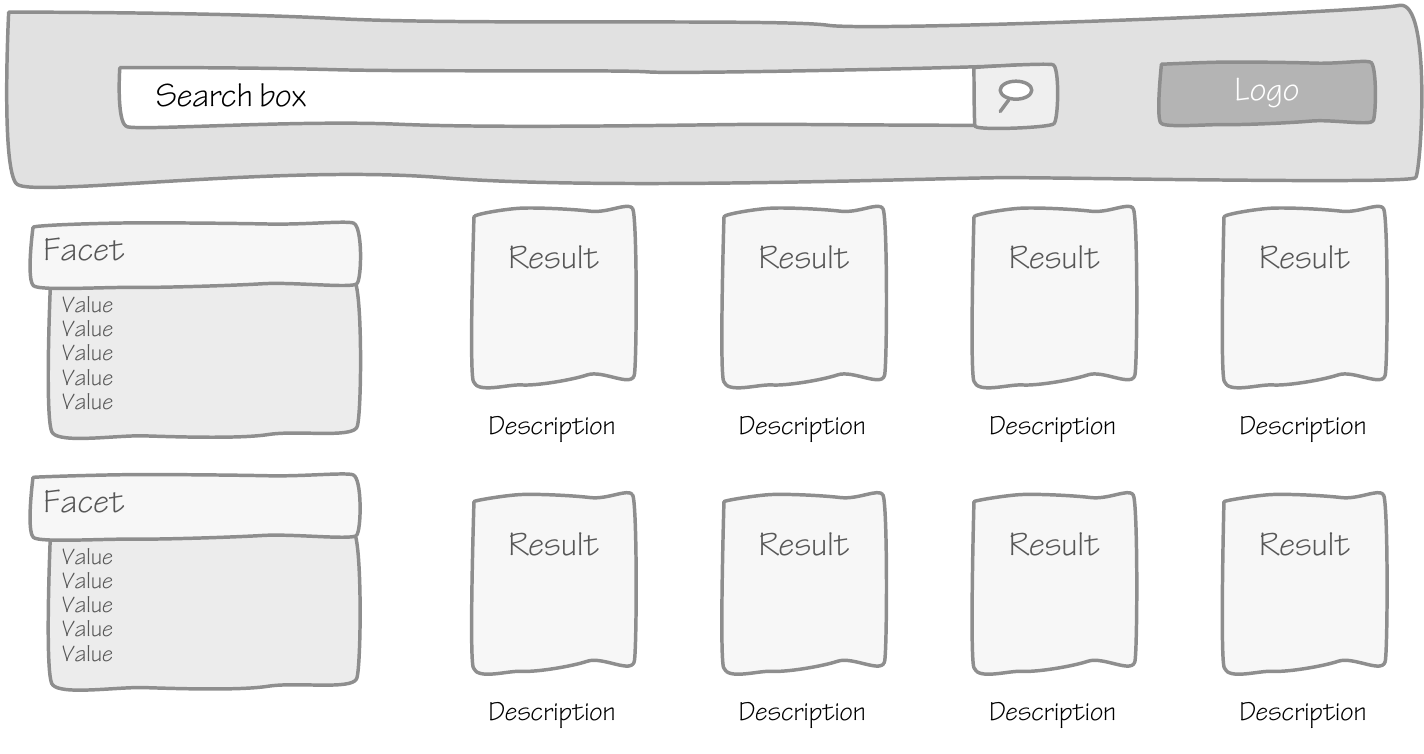}
\caption{A product search layout.}
\label{fig:productsearch}
\end{subfigure}
\\
\begin{subfigure}{0.4\textwidth}
\includegraphics[width=\textwidth]{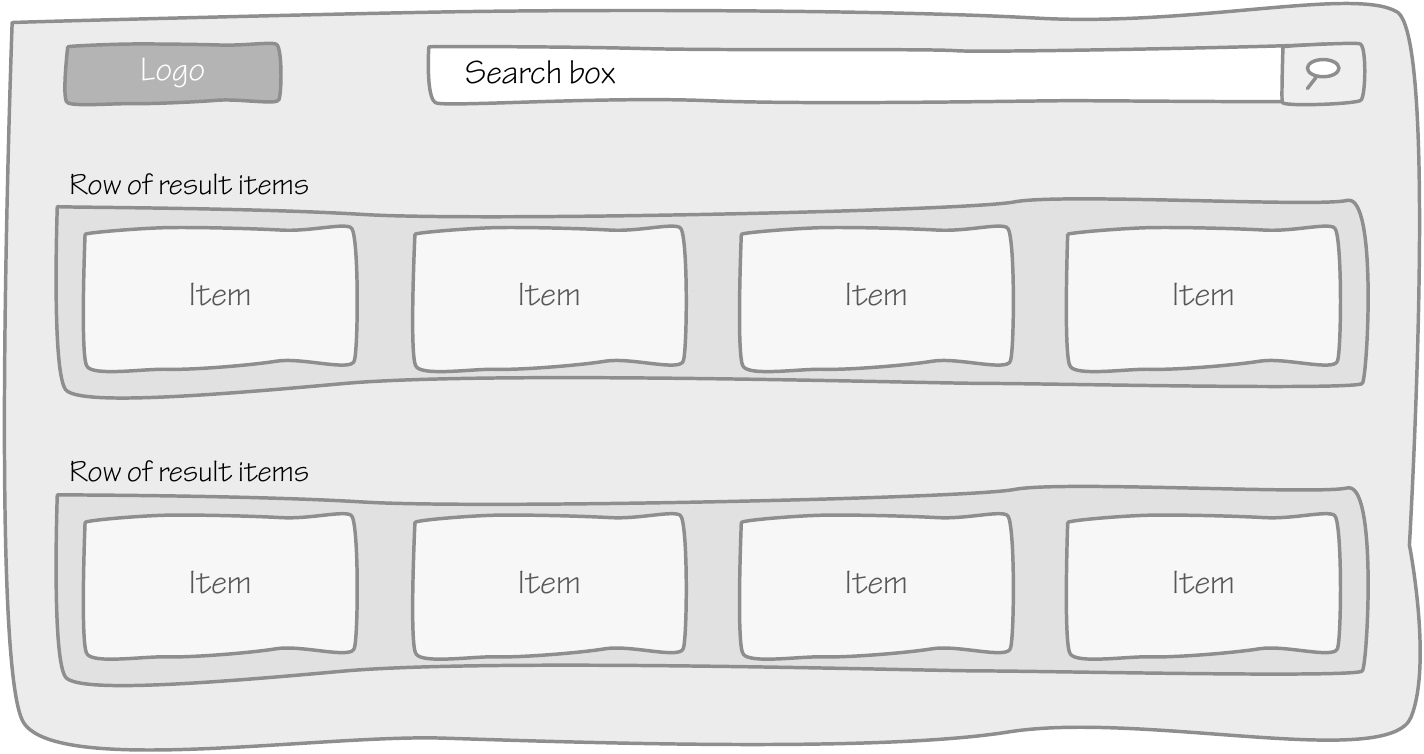}
\caption{A video recommendation layout.}
\label{fig:videosearch}
\end{subfigure}
\hspace*{1.5cm}
\begin{subfigure}{0.4\textwidth}
\includegraphics[width=\textwidth]{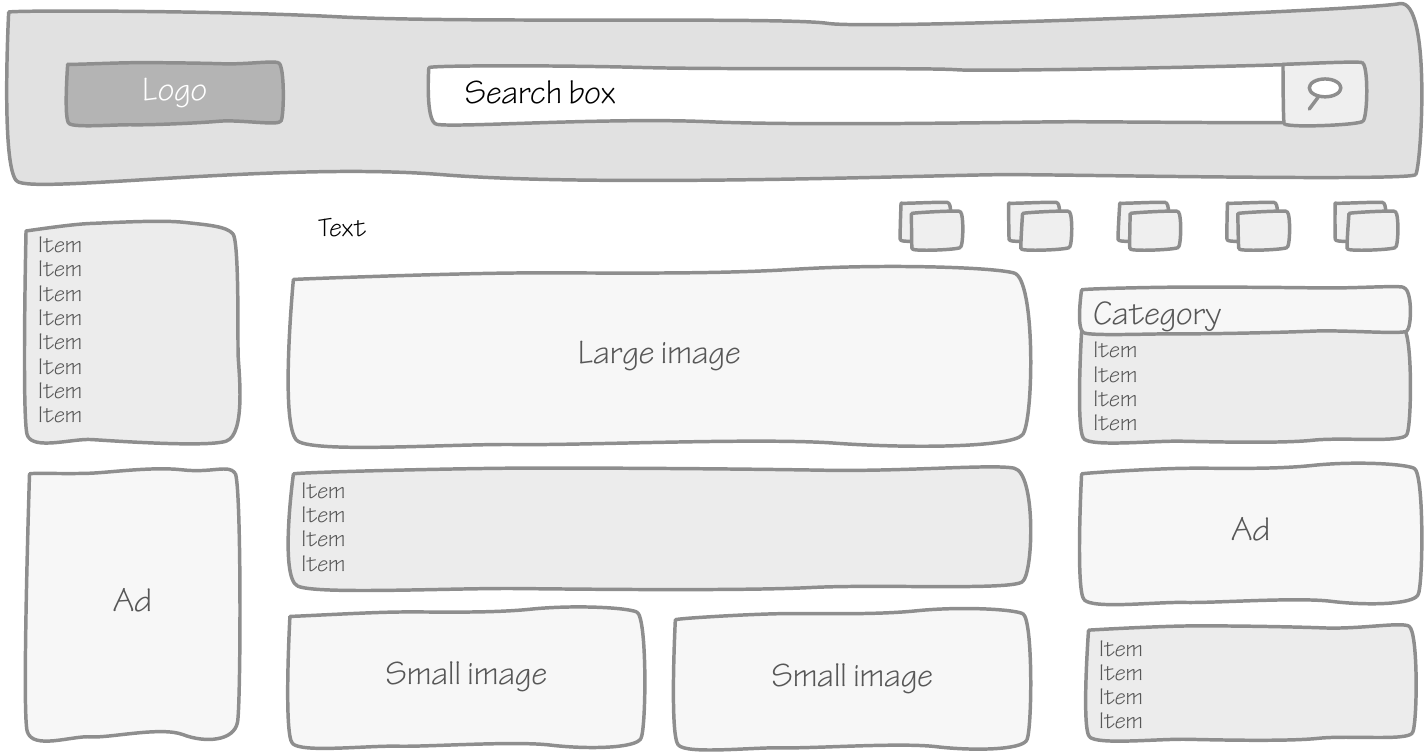}
\caption{An advertisement layout.}
\label{fig:advertisements}
\end{subfigure}
\caption{Examples of different ranking settings and result presentations.}
%%% don't remove; weird bug in counter
\addtocounter{figure}{-1}
\label{fig:complexexamples}
\end{figure*}

The examples of result presentation layouts given in Fig.~\ref{fig:complexexamples} are by no means an exhaustive list; moreover, new layouts and search settings continue to be introduced.
The examples show that in some cases the top-down bias assumption of traditional \ac{LTR} is misguided.
Consequently, if there is a mismatch between the assumed preferred display order and the users' preferences, the expected user experience will be degraded. 
If the wrong display order is assumed then even a ranker that perfectly predicts the relevance order may not display the most relevant document on the ideal position.
We call settings where the ideal display order is unclear a priori \emph{complex ranking settings}, because they involve a double algorithmic problem: both traditional relevance ranking as well as finding the best way to position the ranking.
Thus, in these settings the task is both to infer the user preferred relevance order -- what documents does the user want to see? -- as well as the user preferred display order -- in what order will the user consider the display positions?
User behavior studies can help address this problem; for instance, the eye tracking studies listed above can reveal how users examine a layout and thus infer the preferred display order. 
Such user studies are expensive to perform at scale and it may be hard to find participants that represent the  target user base well. 
Moreover, since layouts change often the results of previous studies may rapidly become obsolete.

As an alternative we investigate whether ranking in complex ranking settings can be learned from user interactions. 
We propose to use Deep \ac{RL}, to infer a display order that best satisfies users, e.g., to infer that the most relevant document is to be displayed in the top-center in the case of a result layout as in Fig.~\ref{fig:videosearch}. 
We introduce two approaches: \begin{inparaenum}[(1)] \item a standard GRU  \cite{cho2014learning}  that implicitly learns the preferred display order, and \item a novel method that explicitly models a preference in display positions\end{inparaenum}. 
We call the latter method the \acfi{DRM} since it learns how to rank the display positions and with which documents to fill those positions.
Our experiments on three large publicly available \ac{LTR} datasets show that only \OurModel{} is capable of successfully addressing complex ranking settings while all other methods are unable to break from the top-down assumption. 
As a result, \OurModel{} achieves the highest retrieval performance in all of the cases where the traditional top-down assumption does not hold while matching the state-of-the-art when the assumption does hold.

In this paper the following research questions are addressed:
 \begin{enumerate}[label={\bf RQ\arabic*},leftmargin=*,nosep]
    \item Can existing ranking methods handle complex ranking settings? \label{rq:existing}
    \item Does \OurModel{} outperform existing methods in complex ranking settings? \label{rq:drmbetter}
%    \item Can \OurModel{} learn complex rankings from SERP-level feedback? \label{rq:serpreward}
\end{enumerate}
%
%Section~\ref{sec:complexrankingproblem} provides a formal definition of the complex ranking problem.
%Section~\ref{sec:background} provides background material on \ac{RL} methods related to the baseline method we introduce in Section~\ref{sec:method} and the novel method in Section~\ref{sec:novelmethod}. 
%Section~\ref{sec:experiments} describes the experimental setup we use to simulate complex ranking settings, and Section~\ref{sec:results} discusses the results of our experiments.
%Section~\ref{sec:related} will discuss related work in non-traditional \ac{LTR} settings and learning in an online setting.
%Section~\ref{sec:conclusion} concludes the paper.

%\begin{figure}[tb]
%\centering
%%\vspace{-.4\baselineskip}
%\includegraphics[width=0.6\columnwidth]{img/introduction/tenbluelinks}
%\caption{Example of a traditional \emph{ten blue links} display.}
%\label{fig:tenbluelinks}
%%\vspace*{-2\baselineskip}
%\end{figure}
% !TEX root = sigir2018-complex-rankings.tex

\section{Problem Setting}
\label{sec:complexrankingproblem}

The idea of a complex ranking setting was introduced in Section~\ref{sec:intro}; this section will formalize the concept and explain why traditional \ac{LTR} methods are unable to deal with them.

\subsection{Ranking settings}

The core task in \ac{LTR} is to order a set of documents to maximize the ease by which a user can perform their search task. Here we assume the user has a preferred order in which they consider the available documents $D = \{ d_1, \ldots d_n \}$. 
This order depends only on the documents and the user's search task and thus we call this the preferred \emph{relevance order}. 
Furthermore, documents are displayed to the user in a layout that has several positions for displaying them. We will assume that a layout has $k$ display positions $P=\langle p_1, \ldots, p_k\rangle$, at each of which a single document can be presented. 
The user considers the display positions one by one, and is expected to do this in a certain order, i.e., subject to a certain position bias \cite{joachims2007evaluating}.
We interpret this order in display positions as the user's preferred \emph{display order}.
Thus, in order to provide the optimal user experience, the most relevant document should be displayed in the position considered first, and so forth.
In other words, documents must be displayed to the user so that there is a correspondence between the preferred relevance order and the display order.
We call such a combination of a preferred relevance order and a preferred display order a \emph{ranking setting}, and formally define it as follows:

\newcommand{\PositionPreference}{>_\mathit{pos}}
\newcommand{\DocumentPreference}{>_\mathit{doc}}

\begin{definition}
A \emph{ranking setting} consists of a set of documents $D = \{d_1, \ldots, d_n\}$, an ordered set of positions $P = \langle p_1, \ldots, p_k\rangle$ as well as user preferences for positions ($\PositionPreference$) and for documents ($\DocumentPreference$).
The user has a preferred \emph{relevance order}, so that all documents can be ordered according to their preference:
\begin{align}
d_i \DocumentPreference d_j \DocumentPreference \cdots \DocumentPreference d_l.
\end{align}
Similarly, positions can be ordered to their preferred \emph{display order}:
\begin{align}
p_i \PositionPreference p_j \PositionPreference \cdots \PositionPreference p_l.
\end{align}
Lastly, a ranking setting has an \emph{ideal ranking} $R = [d_i, \ldots, d_j]$, this is an ordering of the $k$ most preferred documents aligned with preference between display positions:
\begin{align}
\forall 0 < i, j \leq k, \quad R_i \DocumentPreference R_j \leftrightarrow p_i \PositionPreference p_j.
\end{align}
Thus, $R$ places the $i$-th most preferred document on the $i$-th most preferred position, for the top $k$ documents.
Note that the ideal ranking $R$ is not necessarily the same as the preferred relevance order. While the most relevant document is always on the first position in the relevance order, its position in $R$ depends on the preferred display order.
\end{definition}

\subsection{Simple ranking settings}
\label{sec:simplesetting}
We define \emph{simple} and \emph{complex} ranking settings based on whether the display order is known a priori. 
In cases where it is known, the mismatch between the relevance order and the display order can always be avoided:

\begin{definition}
A \emph{simple} ranking setting is a ranking setting where the user's preferred display order is known a priori.
As a result, the ordered set of display positions $P = \langle p_1, \ldots, p_k\rangle$ can be ordered to the users preferences so that:
\begin{align}
\forall 0 < i, j \leq k, \quad i < j \leftrightarrow p_i \PositionPreference p_j.
\end{align}
Consequently, the ideal ranking $R$ is then always aligned with the user preferred relevance order. Thus, if the document set $D$ is extended with one document: $D' = D \cup \{ d' \}$, then in a simple ranking setting the ideal ranking for $D'$, $R^{D'}$, does not change the relative ordering of documents in $R^{D}$.
Therefore, in a simple ranking setting the ideal ranking of $D'$ is always the ideal ranking of $D$ with $d'$ inserted at some position:
\begin{align}
R^{D'} = [R^{D}_1, R^{D}_2, \ldots,  R^{D}_i, d',  R^{D}_{i+1}, R^{D}_{i+2}, \ldots ].
\end{align}
\end{definition}

\noindent%
The \emph{ten blue links} layout is an example of a simple ranking setting, where the top-down display order is well studied \cite{hotchkiss-eye-2005}. Correspondingly, \ac{LTR} has been very effective in this setting and has focused on finding the user's preferred relevance order.
Most \ac{LTR} methods are based on functions that score documents independently and then rank documents according to their scores. This works well in simple ranking settings.
For instance, consider four documents with the preferences:
\begin{align}
d_1 \DocumentPreference d_2 \DocumentPreference d_3 \DocumentPreference d_4,
\end{align}
and three display positions with the preferences:
\begin{align}
p_1 \PositionPreference p_2 \PositionPreference p_3.
\label{eq:simpledisplaypreference}
\end{align}
The two sets $D = \{d_1, d_3, d_4\}$ and $D' = \{d_1, d_2, d_3, d_4\}$ have the following ideal rankings:
\begin{align}
R^{D} = [d_1, d_3, d_4],\\
R^{D'} = [d_1, d_2, d_3].
\end{align}
Thus any scoring function $f$ that correlates with the relevance order in the sense that $f(d_1) > f(d_2) > f(d_3) > f(d_4)$, will provide the ideal rankings for both cases. 
While this is usually not acknowledged explicitly, \ac{LTR} methods have assumed that the ideal ranking should be aligned with the user's preferred relevance order.

\subsection{Complex ranking settings}

Section~\ref{sec:intro} discussed several prevalent settings where the preferred display order is unclear. In these cases some performance is lost by the mismatch between the relevance order and the display order, as even when the most relevant documents are found they may not be displayed in the most efficient manner.
We define the complex ranking setting by whether the display order is known a priori:

\begin{definition}
A ranking setting where the user preferred display order is unknown a priori is called a \emph{complex} ranking setting. 
Therefore, for the document sets $D$ and $D' = D \cup \{ d' \}$, the ideal ranking of $D'$, $R^{D'}$, may change the relative ordering of documents in $D$ with respect to the ideal ranking of $D$, $R^{D}$, if the ranking setting is \emph{complex}.
\end{definition}

\noindent%
If we consider the same example as in Section~\ref{sec:simplesetting} but change the display order from \eqref{eq:simpledisplaypreference} to
\begin{align}
p_2 \PositionPreference p_1 \PositionPreference p_3,
\end{align}
then the ideal rankings for $D = \{d_1, d_3, d_4\}$ and $D' = D \cup \{d_2\}$ are:
\begin{align}
R^{D} = [d_3, d_1, d_4],\\
R^{D'} = [d_2, d_1, d_3].
\end{align}
Unlike the simple ranking setting discussed in Section~\ref{sec:simplesetting}, there is no function that can score documents independently and provide both ideal rankings: since such a function should provide $f(d_3) > f(d_1)$ in the case of $D$ and $f(d_3) < f(d_1)$ in the case of $D'$. 
A possible solution would be a binary document scoring function that considers entire document set as one of its arguments, $f(d, D)$, thus allowing it to distinguish between $D$ and $D'$. 
However, the computational costs of such a model that drops the document independence assumption would make it impossible to scale to any practical usage~\citep{liu2009learning}.

Instead, in Section~\ref{sec:novelmethod} we propose another approach, one where the model sequentially selects a document and a position at which to place it. 
Thus, in contrast with \ac{LTR} in the simple ranking setting, this method simultaneously learns both the relevance order and the display order. 

% !TEX root = sigir2018-complex-rankings.tex

\section{Background}
\label{sec:background}

%In this paper an \ac{RL}  approach will be applied to the complex ranking setting.
This section describes the \ac{RL} concepts related to the novel methods that are introduced in Section~\ref{sec:method} and \ref{sec:novelmethod}.

\subsection{Markov decision processes}
\label{sec:backgroundmdp}

\ac{RL} methods are used to solve decision problem that can be modelled as \acp{MDP} \cite{sutton1998reinforcement}.
An \ac{MDP} models a sequential decision process where an agent executes a chosen action at each time-step; the action will then affect the state of the world in the subsequent time-step.
An \ac{MDP} model consists of states $S$, actions $A$, transitions $T$, rewards $R$, and policies $\pi$. 
The \textbf{states} $S$ represents the set all possible states the process may be in at any time-step. %, a state $s_t$ is assumed to contain all the information required to make a decision at time-step $t$.
The \textbf{actions} $A$ are the set of actions an agent can take at every time-step; this choice is the only part of the \ac{MDP} the agent has direct control over.
The \textbf{transitions}: after each time-step the state changes according to the action taken. These transitions are represented by the distribution $T(s_{t+1} \mid s_t, a_t)$.
The \textbf{rewards}: the reward function $R$ provides the utility of taking an action in a state: $R(s,a)$.
The \textbf{policy}: the actions of the agent are drawn from the policy distribution: $\pi(a\mid s)$. 

The goal of an \ac{RL} method is to find the policy that maximizes the expected (discounted) reward in future time steps:
\begin{align}
\begin{split}
\mathbbm{E}[R(s_t, a_t) & + \gamma R(s_{t+1}, a_{t+1}) \\
& + \gamma^2 R(s_{t+2}, a_{t+2}) + \ldots \mid S, A, T, R, \pi],
\end{split}
\end{align}
where $0 \leq \gamma \leq 1$ acts as a discount factor. 
Thus, the optimal policy should account for both the immediate reward it receives from an action but also the rewards it expects to receive from subsequent steps. 
This means that planning is involved in the decision process, which fits well with complex ranking settings where a relevant document should not always be placed on the first position.

\subsection{Model-free reinforcement learning}

The methods we introduce in Section~\ref{sec:method}~and~\ref{sec:novelmethod} are based on the \ac{DQN} \cite{DQN2015}.
\ac{DQN} is a model free \ac{RL} method; such methods do not explicitly model the transitions or rewards of the \ac{MDP}. 
Instead they work with Q-values, which represent the expected value of a state-action pair: $Q(s, a)$ \cite{sutton1998reinforcement}. 
The value of a state-action pair is the immediate reward and the expected discounted future reward:
\begin{align}
Q(s, a ) &= R(s, a) + \gamma \sum_{s'} T(s' \mid s, a) \max_{a'} Q(s',a').
\end{align}
Thus, the Q-value of a state-action pair depends on the expected value of the next state. 
Here, the maximum over the actions is taken since the agent can be optimistic about its own action selection.
To allow for generalization, in deep Q-learning these values are estimated by a model. Thus the weights $\theta$ should be found so that:
\begin{align}
Q(s, a, \theta) &= R(s, a) + \gamma \sum_{s'} T(s' \mid s, a) \max_{a'}  Q(s',a', \theta).
\end{align}
Furthermore, \ac{DQN} learns these values from experience, i.e., by executing its policy and recording its experience.
At each time-step this gives a transition pair of the state $s$, action $a$,  observed next state $s'$, and observed reward $r$: $\langle s, a, r, s'\rangle$. 
Then $\theta$ can be updated towards:
\begin{align}
Q(s, a, \theta) &= r + \gamma \max_{a'} Q(s', a', \theta).
\end{align}
Over time the Q-values will converge on the true values with the transitions $T$ and reward function $R$ implicitly learned.
%Because the latest experiences are sequentially dependent
%This should not be done directly from the latest experience since the transitions are sequentially dependent. 
\ac{DQN} uses an experience replay buffer \cite{lin1992self} that keeps a large number of past transitions and updates on a batch sampled from this buffer. This mitigates the sequential dependency between recent experiences.

Another issue is that the model is updated towards a target that is predicted from its own weights, making the process very unstable. As a solution, \ac{DQN} uses an older version of the model to predict future the Q-values from:
\begin{align}
Q(s, a, \theta_{T}) &= r + \gamma \max_{a'} Q(s', a', \theta_{L}),
\end{align}
where $\theta_{T}$ is being trained and $\theta_{L}$ is a previous set of weights. 
Periodically the $\theta_{T}$ is transferred to $\theta_{L}$: $\theta_{L} \gets \theta_{T}$.
As a result, the weights are updated to a target that remains stationary for long periods of time, stabilizing the learning process. 
Lastly, \ac{DQN} is prone to over estimate certain Q-values, because the max operation is biased towards actions that are erroneously overestimated. 
As a solution, \citet{van2016deep} proposed Double \ac{DQN}, where the maximizing action is estimated on $\theta_{T}$:
\begin{align}
a' &= \argmax_a Q(s', a, \theta_{T})\\
Q(s, a, \theta_{T}) &= r + \gamma Q(s', a', \theta_{L}).
\end{align}
Thus, an overestimated value in $Q(s, a, \theta_{L})$ is less likely to be selected, since $\theta_{T}$ and $\theta_{L}$ are somewhat independent.

\begin{algorithm}[t]
\caption{Double DQN \cite{DQN2015, van2016deep} a Model-Free \ac{RL} method.} 
\label{alg:dqn}
\begin{algorithmic}[1]
\STATE \textbf{Input}: weights: $\theta_0$, replay size: $M$, transfer steps: $N$
\STATE $E \leftarrow []$ \hfill \textit{\small // initialize experience replay} \label{alg:line:initexp}
\STATE $\theta_L \leftarrow \theta_0$ \hfill \textit{\small // initialize label network} 
\FOR{$i \in [1, 2, \ldots ,]$}
    \WHILE{$|E| < M$}
    	\STATE $\langle s,a,r,s'\rangle \leftarrow \textit{execute\_policy}(\theta_i)$ \label{alg:line:executepolicy}
         \STATE $E \leftarrow append(E, \langle s,a,s',r\rangle)$  
    \ENDWHILE
    \STATE $\mathcal{L} \leftarrow 0$ \hfill \textit{\small // initialize loss function}  \label{alg:line:initloss}
    \FOR{$j \in [1, \ldots, \textit{batch\_size}]$}
        \STATE $\langle s,a,r,s'\rangle \leftarrow \textit{sample}(E)$  \hfill \textit{\small // sample without replacement} \label{alg:line:samplereplay}
        \STATE $a' \leftarrow \argmax_a Q(s', a, \theta_i)$ \label{alg:line:maxaction}
       \STATE $\mathcal{L} \leftarrow \mathcal{L} + (Q(s, a, \theta_i) - r - \gamma Q(s', a', \theta_L))^2$ \label{alg:line:regressionloss}
    \ENDFOR
    \STATE $\theta_{i+1} = \textit{gradient\_descent\_update}(\mathcal{L})$  \label{alg:line:update}
    \IF{$i \mod N = 0$}
        \STATE $\theta_L = \theta_i$  \label{alg:line:transfer}
    \ENDIF
\ENDFOR
\end{algorithmic}
\end{algorithm}

The full Double DQN procedure is displayed in Algorithm~\ref{alg:dqn}. 
At first, the Experience Replay buffer is empty (Line~\ref{alg:line:initexp}). 
Then the initial policy is executed until the buffer is filled to contain $M$ past transitions (Line~\ref{alg:line:executepolicy}). 
 %When the buffer is of the required size the loss can be calculated over a sampled batch. 
Subsequently, the loss $\mathcal{L}$ is initialized (Line~\ref{alg:line:initloss}) and several transitions are sampled from the replay buffer (Line~\ref{alg:line:samplereplay}). 
The maximizing action according to the train-network $\theta_i$ is determined (Line~\ref{alg:line:maxaction}), and the loss is updated with the regression loss for the current Q-value according to the train-network and the estimated value according to the label-network:
\begin{align}
(Q(s, a, \theta_i) - r - \gamma Q(s', a', \theta_L))^2.
\end{align}
The weights are updated according to the calculated loss (Line~\ref{alg:line:update}) providing a new model $\theta_{i+1}$. 
After $N$ update steps the label network is replaced with a copy of the current training network (Line~\ref{alg:line:transfer}). 
%This allows the Q-values to propagate while still having stable optimization.
When the Q-values have been learned, the estimated optimal policy is derived by taking the maximizing action for each state.

% !TEX root = sigir2018-complex-rankings.tex

\section{Reinforcement Learning to Rank}
\label{sec:method}

This section introduces a baseline approach to the complex ranking setting. First Section~\ref{sec:rankmdp} discusses how \ac{LTR} can be approached as an \ac{RL} problem, then Section~\ref{sec:standardmethod} introduces a baseline approach to the problem. %In contrast, Section~\ref{sec:novelmethod} will describe our novel method that is designed specifically for the Complex Ranking setting.

\begin{algorithm}[tb]
\caption{Sampling an episode with the baseline \ac{GRU} model.} 
\label{alg:grupolicy}
\begin{algorithmic}[1]
%\STATE \textbf{Input}: Model $f$, Expl. Degree $\epsilon$, Weights: $\theta$, SERP size: $k$
\STATE $\mathbf{D}_q  \leftarrow \textit{receive\_query}$  \hfill \textit{\small// query and pre-selection of doc.}
\STATE $\mathbf{h}_0 \leftarrow \mathbf{0}$ \hfill \textit{\small// initialize hidden state}
\STATE $\mathbf{R} \leftarrow []$ \label{line:gru:dinit}
\FOR{$t \leftarrow 1 \ldots  k$ }
    \IF{coinflip with $\epsilon$ probability}
        \STATE $\mathbf{d}_t \leftarrow \textit{sample\_document}(\mathbf{D}_q)$ \hfill \textit{\small// explorative action}  \label{line:gru:explore}
    \ELSE
        \STATE $\mathbf{q} \leftarrow [0, \ldots , 0] $ \hfill \textit{\small// initialize zero document score vector}
        \FOR{$d_i \in \mathbf{D}_q$ }
           \STATE $\mathbf{h}' \leftarrow GRU(\mathbf{h}_{t-1}, \mathbf{d}_i)$
            \STATE $\mathbf{q}_i \leftarrow Q(h', d_i, \theta)$ \hfill \textit{\small// estimate Q-value for doc. (Eq~\ref{eq:standardqvalue})} \label{line:gru:qvalue}
        \ENDFOR
        \STATE $i \leftarrow \argmax{\mathbf{q}}$ \hfill \textit{\small// select doc. with highest q-value}
        \STATE $\mathbf{R}_t \leftarrow \textit{gather}(\mathbf{D}_q, d_i)$ \hfill \textit{\small// add doc. to SERP}
    \ENDIF
    \STATE $\mathbf{D}_q \gets \textit{remove}(\mathbf{D}_q, \mathbf{d}_t)$  \hfill \textit{\small// remove to prevent duplicates} \label{line:gru:removedoc}
    \STATE $\mathbf{h}_t = GRU(\mathbf{h}_{t-1}, \mathbf{d}_t)$ \hfill \textit{\small// update (partial) SERP embedding} \label{line:gru:adddoc}
\ENDFOR
\RETURN $\mathbf{R}$
\end{algorithmic}
\end{algorithm}

\subsection{Ranking as a Markov decision process}
\label{sec:rankmdp}

\ac{LTR} has been approached as an \ac{RL} problem in the past~\cite{wei2017reinforcement, xia2017adapting}; notably \citeauthor{xia2017adapting}~\cite{xia2017adapting} used a policy gradient method for search result diversification. 
The methods introduced in this paper are based on Q-learning instead of policy gradients, however they also approach ranking as an \ac{MDP}.

The complex ranking setting is defined by the user preferred display order being unknown a priori. As a result, it is impossible to acquire labelled data, e.g., from human annotators.
As an alternative, we will use \ac{RL} to learn from user interactions.
First, the ranking problem must be formalized as an \ac{MDP}; as described in Section~\ref{sec:backgroundmdp}, this means that states $S$, actions $A$, transitions $T$ and rewards $R$ have to be specified.
In this paper, we approach ranking as a sequential decision problem, where every document in a ranking is seen as a separate decision.
Accordingly, the \textbf{states} $S$ encode the query information and all possible partial rankings; the initial state $s_0$ represents the query and an empty ranking.
The \textbf{actions} $A$ consist of the available documents $D_q$, with the exception of documents already in $s_t$ to prevent duplicate placement.
The \textbf{transitions} $T$ are deterministic between the partial rankings in $S$: adding a document $d'$ to $s_t = (q, [d_1,\ldots,d_i])$ transitions to $s_{t+1} = (q, [d_1,\ldots,d_i, d'])$ exclusively.
Since no ranking in $S$ exceeds the number of display positions $k$, every episode lasts $k$ steps and any state $s_t$ at time step $t=k$ is an end-state.
Lastly, the \textbf{rewards} $R$ can be given at the document-level or at the SERP-level; for a chosen $\textit{discount}$ function, the document level reward is given by: 
\begin{align}
R_\mathit{doc}(s_t, a_t) =  \frac{2^{\mathit{rel}(d_{a_t}) - 1}}{\mathit{discount}(t)}. 
\label{eq:grudocreward}
\end{align}
The discount function can be chosen to match the \ac{DCG}: $\mathit{discount}(i) = \log_2(i + 1)$.
However, other discount functions can simulate different complex ranking settings.
The SERP-level reward is only given for actions that complete rankings and simply sums the document-rewards in the ranking:
\begin{align}
R_\mathit{SERP}(s_t,a_t) =
\begin{cases}
0, & t < k\\
\sum^k_{i=1} R_\mathit{doc}(s_i,a_i), & t = k.\\
\end{cases}
\end{align}
The SERP-level reward simulates challenging settings where the reward cannot be broken down to the document level.
Since every episode is limited to $k$ steps, future rewards are not discounted, that is, $\gamma = 1$.
So the aim of the policy is to maximize the expected reward over the entire episode.

\subsection{A baseline approach to complex ranking}
\label{sec:standardmethod}

In order to use Deep Q-Learning on the complex ranking setting, a model that can estimate Q-values is required.
This section will introduce a baseline estimator, before Section~\ref{sec:novelmethod} introduces a novel model specialized for the complex ranking setting.

\newcommand{\terug}{\mbox{}\hspace*{-2.25mm}}

Since ranking is approached as a sequential process, a \ac{RNN} is used to encode the ranking so far and estimate the value of adding the next document.
Unlike previous work~\cite{wei2017reinforcement, xia2017adapting}, we use a \ac{GRU}~\cite{cho2014learning} instead of a plain \ac{RNN}.
The benefit of a \ac{GRU} is that, similar to a \ac{LSTM} \cite{hochreiter1997long}, it has a form of explicit memory, allowing it to \emph{remember} values over multiple iterations in the sequence.
Compared to the \ac{LSTM} model a \ac{GRU} has fewer parameters since it only has an \emph{update} gate vector $z_t$ and \emph{reset} gate vector $r_t$.
The \ac{GRU} model can be formulated as follows:
\begin{align}
\terug z_t &= \sigma(W_z x_t + U_z h_{t-1} + b_z), \\
\terug r_t &= \sigma(W_r x_t + U_r h_{t-1} + b_r), \\
\terug h_t &= GRU(h_{t-1}, x_t) \\ &= z_t \circ h_{t-1} + (1-z_t) \circ \tanh(W_h h_{t-1} + U_h(r_t \circ h_{t-1}) + b_h)\terug
\end{align}
where $\circ$ is the Hadamard product and the matrices $W$, $U$ and vectors $b$ are the weights to be optimized.

It would make sense to start the Q-value estimation by encoding the query $\mathbf{q}$.
Unfortunately, no query-level features are available in public \ac{LTR} datasets (Section~\ref{sec:experiment:dataset}), thus we use zero initialization for the hidden state:
\begin{align}
h_0 &= \mathbf{0}.
\end{align}
Then, for all the documents pre-selected for the query $D_q$ an embedding is made, with $\sigma$ as the \ac{ReLU} function and document feature vector $\mathbf{d}$:
\begin{align}
\mathbf{\widehat{d}} = \sigma(W_d\mathbf{d} + b_d).
\end{align}
These document representations are shared for each step in the process; if at time-step $t$ the document $d$ is selected the Q-value is calculated as follows:
\begin{align}
h_t &= GRU(h_{t-1},\mathbf{\widehat{d}}), \\
Q(s_t, d_t, \theta) &= v_q^T\sigma(W_qh_t + b_q) + u_q, \label{eq:standardqvalue}
\end{align}
where the matrix $W_q$, the vectors $v_q$, $b_q$ and scalar $u_q$ are weights to be optimized.
%
%Based on the Q-values a policy can be derived, the \emph{exploitive} policy would always perform the action $a_t$ that maximizes $Q(s_t, a_t, \theta)$. However, during learning it is important to \emph{explore} by trying out actions that are estimated to be suboptimal but may actually be very beneficial. We applied 
%

For clarity, Algorithm~\ref{alg:grupolicy} describes the policy in detail; an epsilon greedy approach is used to account for exploration, meaning that at every time step a random action is performed with probability $\epsilon$.
Initially the ranking $\mathbf{R}$ is empty (Line~\ref{line:gru:dinit}).
For $k$ iterations with probability $\epsilon$ a document is uniformly sampled (Line~\ref{line:gru:explore}). Otherwise, the Q-value for adding a document is computed for every available document (Line~\ref{line:gru:qvalue}).
The document with the highest Q-value is added to the SERP (Line~\ref{line:gru:adddoc}) and removed from the set of available documents (Line~\ref{line:gru:removedoc}); this prevents it from appearing twice in $\mathbf{R}$.
Finally, after $\mathbf{R}$ is completed, it is used by DQN (Algorithm~\ref{alg:dqn}); DQN interprets every document placement as an action as described in Section~\ref{sec:rankmdp}.

Since this approach sequentially decides whether to place documents, it could learn a policy that does not place the most relevant document first.
Potentially, it could wait until the best display position before placing it.
The risk here is that if it saves more relevant documents than positions left to fill, some of them will not be displayed.

\begin{algorithm}[tb]
\caption{Sampling an episode with \ac{DRM}.} 
\label{alg:drmepisode}
\begin{algorithmic}[1]
%\STATE \textbf{Input}: Model $f$, Expl. Degree $\epsilon$, Weights: $\theta$, SERP size: $k$
\STATE $\mathbf{D}_q  \leftarrow \textit{receive\_query}$  \hfill \textit{\small// query and pre-selection of doc.}
\STATE $\mathbf{P} \leftarrow [p_1, \ldots ,p_ k]$ \hfill \textit{\small// available positions}
\STATE $h_0 \leftarrow \mathbf{0}$ \hfill \textit{\small// initialize hidden state}
\STATE $\mathbf{R}, \mathbf{I} \leftarrow [], []$ \hfill \textit{\small// initialize ranking and selected positions} \label{line:drm:init}
\FOR{$t \leftarrow 1 \ldots  k$ }
    \IF{coinflip with $\epsilon$ probability}
        \STATE $\mathbf{d}_t \leftarrow \textit{sample\_document}(\mathbf{D}_q)$ \hfill \textit{\small// explorative doc. action} \label{line:drm:docexplore}
    \ELSE
        \STATE  \textit{\small// find the document with highest Q-value (Eq.~\ref{eq:drm:docvalue})}
        \STATE $\mathbf{R}_t \leftarrow \argmax_{d_i\in \mathbf{D}_q}{Q(h_{t-1}, d_i, \theta)}$ \label{line:drm:docexploit}
    \ENDIF
    \STATE $\mathbf{D}_q \leftarrow \textit{remove}(\mathbf{D}_q, \mathbf{R}_t)$  \hfill \textit{\small// remove to prevent duplicates} \label{line:drm:remove}
    \IF{coinflip with $\epsilon$ probability}
        \STATE $\mathbf{p}_t \leftarrow \textit{sample\_position}(\mathbf{P})$ \hfill \textit{\small// explorative pos. action} \label{line:drm:posexplore}
    \ELSE
         \STATE  \textit{\small// find the position with highest Q-value (Eq.~\ref{eq:drm:posvalue})}
        \STATE $\mathbf{P}_t \leftarrow \argmax_{p_i\in \mathbf{P}}{Q(h_{t-1}, \mathbf{R}_t, p_i, \theta)}$ \label{line:drm:posexploit}
    \ENDIF
    \STATE $\mathbf{P} \leftarrow \textit{remove}(\mathbf{P}, \mathbf{p}_t)$  \hfill \textit{\small// make position unavailable} \label{line:drm:posremove}
    \STATE $\mathbf{h}_t = GRU(\mathbf{h}_{t-1}, \mathbf{d}_t, \mathbf{p}_t)$ \hfill \textit{\small// update (partial) SERP embedding}
\ENDFOR
\RETURN $\mathbf{R}, \mathbf{I}$
\end{algorithmic}
\end{algorithm}

\section{Double-Rank for Complex Ranking} % Settings}
\label{sec:novelmethod}

Section~\ref{sec:standardmethod} used a standard \ac{GRU} to sequentially choose what documents to place. 
In this section we introduce a method that sequentially chooses a document and then a display position to place it in.  
This approach explicitly models the duality of the complex ranking setting: the preferred relevance order and the preference in display positions. Because it ranks both documents and positions we call it the \acfi{DRM}. 
The key insight behind \ac{DRM} is that by choosing in what order to fill display positions, it can avoid having to \emph{save} a good document for a later position. 
Since it can start placement at the most preferred position. 
\ac{DRM} produces a list of documents $\mathbf{R}$ and a list of positions $\mathbf{I}$; the SERP is then created by placing the documents in $\mathbf{R}$ according to $\mathbf{I}$.
Thus, the selection of $\mathbf{R}$ encapsulates the inferred relevance order and the choice of $\mathbf{I}$ captures the inferred display order.

%The \emph{Complex Ranking} setting is defined by the user having a preference is display positions that is unknown a priori. Thus this setting poses a dual problem: the \emph{traditional} relevance ranking and inferring the preference in display positions. We propose a method that explicitly models the duality of the problem by ranking both the documents as well as the display positions, hence we call it the \acf{DRM}. The insight behind this approach is that if the model can choose in what order to fill display positions, it does not have to \emph{save} a good document for a later position. Since it can start placement at the most preferred position.  Instead of sequentially adding to a list of documents $\mathbf{d}$ \ac{DRM} also creates a list of indices $\mathbf{i}$, the SERP is then created by re-ranking $\mathbf{d}$ according to $\mathbf{i}$. Thus the selection of $\mathbf{d}$ encapsulates relevance ranking and the choice of $\mathbf{i}$ captures the display preference of the user. For this paper, we assume a display has a limited number of $k$ positions, i.e. the number of documents on the first result page.

\subsection{Changing the MDP for \ac{DRM}}
\label{sec:drm:mdp}
The \ac{MDP} as described in Section~\ref{sec:rankmdp} has to be altered to accommodate the \ac{DRM} approach.
Firstly, the \textbf{states} $S$ now include every (partial) ranking of documents $\mathbf{R}$ and every (partial) matching set of positions $\mathbf{I}$.
We arbitrarily choose documents to be selected before their positions, thus at the first state a document is added to $\mathbf{R}$ and at the next state its position is added to $\mathbf{I}$.
%As a result, the states alternate between $\mathbf{R}$ and $\mathbf{I}$ being equal size and $\mathbf{R}$ having one more element.
The \textbf{actions} $A$ also consist of either choosing a document or index depending on the state $s_t$. In other words, if at time step $t$ a document is chosen ($a_t \in D_q$), then at the next step its position is chosen ($a_{t+1} \in P$).
Similarly, the \textbf{transitions} $T$ are deterministic between the states: adding a document $d'$ transitions the state from $s_t = (q, [d_1, \ldots, d_i], [p_1, \ldots ,p_i])$ to $s_{t+1} = (q, [d_1, \ldots, d_i, d'], [p_1, \ldots, p_i])$ exclusively.
Likewise, choosing a position  $p'$ transitions it from $s_t = (q, [d_1,\ldots$, $d_i]$,  $[p_1, \ldots$, $p_{i-1}])$ to $s_{t+1} = (q$, $[d_1,\ldots ,d_i], [p_1,\ldots ,p_{i-1},p'])$.
As a result, every episode now consists of $2k$ steps.
Finally, the \textbf{reward function} has to be adapted slightly.
For the document-level reward this becomes:
 \begin{align}
R_{doc}(s_t, a_t) =
\begin{cases}
0, & a_t \in D_q \\
\frac{2^{\textit{rel}(d_{a_{t-1}}) - 1}}{\textit{discount}(p_{a_{t}})} & a_t \in P, \\
\end{cases}
\end{align}
 where $p_{a_{t}}$ is the selected position at $t$ and $d_{a_{t-1}}$ is the corresponding document selected at the previous step. The discount here only depends on the selected position and not on what time-step it was placed there. The SERP-level reward becomes:
 \begin{align}
R_\mathit{SERP}(s_t,a_t) =
\begin{cases}
0, & t < 2k\\
\sum^{2k}_{i=1} R_{doc}(s_i,a_i), & t = 2k.
\end{cases}
\end{align}
While the reward function is different for the \ac{DRM}, every SERP receives the same total reward as in the baseline \ac{MDP} (Section~\ref{sec:rankmdp}).
%the reward for \ac{DRM} is the same as described in Section~\ref{sec:rankmdp}, meaning that any SERP that comes out of \ac{DRM} receives the same rewards as for the standard \ac{GRU} model.
% Thus re-ordering $\mathbf{R}$ according to $\mathbf{I}$ and then applying Equation~\ref{eq:grudocreward} gives the same total reward.
 
\subsection{The Double-Rank model}

With the \ac{DRM}-\ac{MDP} defined, we can now formulate the model that estimates Q-values and thus learns the ranking policy.
Similar to the baseline model (Section~\ref{sec:standardmethod}), the \ac{DRM} uses a \ac{GRU} network to encode the state.
First the hidden state is initialized:
\begin{align}
h_0 &=\mathbf{0}.
\end{align}
Then an embedding for every document is made, given the document feature vector $\mathbf{d}$:
\begin{align}
\mathbf{\widehat{d}} = \sigma(W_d\mathbf{d} + b_d).
\end{align}
Instead of alternating the input of the \ac{GRU} with representations of documents or positions, we only update the hidden state $h_t$ after every position action.
As an input the embedding of the previously chosen document $\mathbf{\widehat{d}}_{t-1}$ and position $p_{t}$ are concatenated:
\begin{align}
h_t &= GRU(h_{t-2}, [\mathbf{\widehat{d}}_{t-1}, p_{t}]),
\end{align}
where each position $p_t$ is represented by a unique integer.

The Q-values for document actions and position actions are computed differently; both use the concatenation of the last hidden state and the corresponding document embedding: $[h_{t-1}, \mathbf{\widehat{d}}_{t}]$.
For a document action the Q-value is calculated by:
\begin{align}
Q(s_t, d_t, \theta) &= v_q^T\sigma(W_q[h_{t-1}, \mathbf{\widehat{d}}_{t}] + b_q) + u_q, \label{eq:drm:docvalue}
\end{align}
The computation for position action $p_t$ in the subsequent time-step uses the same concatenation (now denoted as $[h_{t-2}, \mathbf{\widehat{d}}_{t-1}]$):
\begin{align}
Q(s_t, p_t, \theta) &= v_{p_t}^T\sigma(W_{p}[h_{t-2}, \mathbf{\widehat{d}}_{t-1}] + b_{p}) + u_{p_t}, \label{eq:drm:posvalue}
\end{align}
where the vectors $v_{p_t}$ and scalar $u_{p_t}$ are unique for the position $p_t$. Since the number of positions is limited, we found it more effective to use unique weights for each of them, in terms of both computational efficiency and learning speed.

The \ac{DRM} policy is displayed in Algorithm~\ref{alg:drmepisode}; the document ranking $\mathbf{R}$ and position ranking $\mathbf{I}$ are initialized (Line~\ref{line:drm:init}). 
Then for $k$ iterations a document and position are selected subsequently. 
First, with an $\epsilon$ probability  a random document is selected as an exploratory action (Line~\ref{line:drm:docexplore}). 
Otherwise, the document with the highest estimated Q-value is selected (Line~\ref{line:drm:docexploit}). It is then removed from the available set $D_q$ to prevent duplicate placement. 
Subsequently, with an $\epsilon$ probability an exploratory position is selected (Line~\ref{line:drm:posexplore}). Otherwise, the position with the highest Q-value is selected (Line~\ref{line:drm:posexploit}); note that this value depends on the previously selected document. 
The position is then made unavailable for subsequently selected documents (Line~\ref{line:drm:posremove}). 
When $k$ documents and positions have been selected, the rankings $\mathbf{R}$ and $\mathbf{I}$ are passed to the \ac{DQN} (Algorithm~\ref{alg:dqn}). 
The user will be shown the documents in $\mathbf{R}$ in the display positions according to $\mathbf{I}$. 
Given a weak reward signal the choice of both  $\mathbf{R}$ and $\mathbf{I}$ can be optimized using \ac{DQN}.

\section{Experimental Setup}
\label{sec:experiments}
This section describes the experiments that were run to answer the research questions posed in Section~\ref{sec:intro}.
%First, we detail the datasets used (Section~\ref{sec:experiment:dataset}), then the reward signals used to simulate different complex ranking settings (Section~\ref{sec:experiment:setting}) and the specific runs that were created (Section~\ref{sec:experiment:runs}).

\subsection{Datasets}
\label{sec:experiment:dataset}
Our experiments are performed over three large publicly available \ac{LTR} datasets \cite{Qin2013Letor, Chapelle2011, dato2016fast}; these are retired validations sets published by large commercial search engines.
Each dataset consists of queries, documents and relevance labels. For every query there is a preselection of documents; while queries are only represented by their identifiers, feature representations and relevance labels are available for every preselected document-query pair. 
Relevance labels range from \emph{not relevant} (0) to \emph{perfectly relevant} (5), and each dataset is divided into training, validation and test partitions.

Firstly, we use the \emph{MSLR-WEB30k} dataset released by Microsoft in 2010~\cite{Qin2013Letor}. 
It consists of 30,000 queries obtained from a retired labelling set of the Bing search engine. 
The dataset uses 136 features to represent its documents; each query has 125 assessed documents on average. 
This dataset is distributed in five folds for cross-fold validation.
Secondly, also in 2010, Yahoo!\ organized a public Learning to Rank Challenge \cite{Chapelle2011} with an accompanying dataset. 
This set consists of 709,877 documents encoded in 700 features and sampled from query logs of the Yahoo! search engine, spanning  29,921 queries.
Thirdly, in 2016 the Istella search engine released a \ac{LTR} dataset to the public \cite{dato2016fast}. 
It is the largest dataset to date with a total of 33,118 queries, an average of 315 documents per query and 220 features. 
Only a training and test partition were made available for this dataset; for parameter tuning we sampled 9,700 queries from the training partition and used them as a validation set.

\subsection{Simulating complex ranking settings}
\label{sec:experiment:setting}

Our experiments are based around a stream of queries, for every received query the system presents the user a SERP generated by the policy (Algorithm~\ref{alg:grupolicy}~or~\ref{alg:drmepisode}). The SERP is displayed to the user in a $k=10$ position display. The system then receives a reward signal from the user, indicative of their satisfaction. These interactions are passed to the DQN (Algorithm~\ref{alg:dqn}) which stores them in the experience replay buffer. When the buffer is sufficiently full, a batch of interactions is sampled and the model is updated accordingly.
The stream of queries is simulated by uniformly sampling queries from the datasets. The reward signal is simulated by the function described in Section~\ref{sec:rankmdp}~or~\ref{sec:drm:mdp}, where the $\mathit{discount}$ function is varied to simulate different complex ranking settings.
All of the $\mathit{discount}$ functions we use are based on \acf{DCG}; we assign every position $p_i$ an integer index and then define:
\begin{align}
\mathit{discount}(p_i) &= \log_2(p_i + 1).
\end{align}
Here, the integer $p_i$ is the index the position has in the user preferred display order. 
Thus, the first observed position will be discounted by $\textit{discount}(1)$, and so on. 
Then we use three display orders as displayed in Table~\ref{tab:displayorders}. 
First, \emph{first-bias} simulates a user who considers positions in the assumed order, resulting in a standard \ac{DCG} reward signal.
Then \emph{center-bias} simulates a user who considers the center position first, e.g., as can happen in a horizontal display. 
Lastly, we consider the \emph{last-bias} preference order, where the last position is considered first, this simulates the worst-case scenario as the display order is the furthest removed from the assumed order.

Note that no method is initially aware of what the display order is; the only indirect indication of the preferred order comes from the received rewards.

\def\cca#1{\cellcolor{black!#10}\ifnum #1>5\color{white}\fi{\bf #1}}
\newcommand{\SingleNr}[1]{\phantom{0}\cca{#1}\phantom{0}}

\begin{table}[tb]
\caption{Three different display order preferences used to simulate different complex ranking settings.
The $p_i$ indicate positions and the numerical values in the cells and the intensity of the color indicate preferred display order. E.g., in the first-bias preference, $p_1$ is the most preferred position and $p_{10}$ the least preferred position.}
\label{tab:displayorders}
\medskip
\centering
%\begin{tabular}{ @{} l c c c c c c c c c c }
%\toprule
% & $p_1$ & $p_2$ & $p_3$ & $p_4$ & $p_5$ & $p_6$ & $p_7$ & $p_8$ & $p_9$ & $p_{10}$ \\
% \midrule
% first-bias & 1 & 2 & 3 & 4 & 5 & 6 & 7 & 8 & 9 & 10 \\
% centre-bias & 9 & 7 & 5 & 3 & 1 & 2 & 4 & 6 & 8 & 10 \\
% last-bias & 10 & 9 & 8 & 7 & 6 & 5 & 4 & 3 & 2 & 1 \\
%\bottomrule
%\end{tabular}
{\setlength\tabcolsep{2.9pt}%
\begin{tabular}{lcccccccccc}
 & $p_1$ & $p_2$ & $p_3$ & $p_4$ & $p_5$ & $p_6$ & $p_7$ & $p_8$ & $p_9$ & $p_{10}$ \\[0.7ex]
 first-bias & \SingleNr{1} & \SingleNr{2} & \SingleNr{3} & \SingleNr{4} & \SingleNr{5} & \SingleNr{6} & \SingleNr{7} & \SingleNr{8} & \SingleNr{9} & \cca{10} \\
 center-bias~~ & \SingleNr{9} & \SingleNr{7} & \SingleNr{5} & \SingleNr{3} & \SingleNr{1} & \SingleNr{2} & \SingleNr{4} & \SingleNr{6} & \SingleNr{8} & \cca{10} \\
 last-bias & \cca{10} & \SingleNr{9} & \SingleNr{8} & \SingleNr{7} & \SingleNr{6} & \SingleNr{5} & \SingleNr{4} & \SingleNr{3} & \SingleNr{2} & \SingleNr{1}\\
\end{tabular}}
\end{table}

For evaluation we use the mean normalized reward, which we call the \ac{P-NDCG}. This is the received reward divided by the maximum possible reward:
\begin{align}
\textit{P-NDCG} = \frac{\sum_{t=1}^k R(s_t,a_t)}{\max_{[a'_0,\ldots, a'_k]} \sum_{t=1}^k R(s_t'a_t')}.
\end{align}
This reward can be seen as NDCG where the discount indices have been permuted.
Note that in the \emph{first-bias} case, \ac{P-NDCG} is the equivalent of regular NDCG.
Lastly, to evaluate the estimated optimal behavior, \ac{P-NDCG} is computed without exploration: $\epsilon = 0$.
%\todo{Please clarify: Though the normalized reward is a better indication of performance, we provide the algorithm the non-normalized reward signal because it is easier to estimate. 
%The \ac{P-NDCG} is calculated for the policies without exploration: $\epsilon = 0$, thus providing the estimated optimal behavior.}

\subsection{Experimental runs}
\label{sec:experiment:runs}

Runs were created with \ac{DRM} (Section~\ref{sec:novelmethod}), the baseline \ac{GRU} (Section~\ref{sec:method}), and the MDP-DIV model~\citep{xia2017adapting}. 
The latter model was introduced for search result diversification; however, by changing the reward function it can be applied to the complex ranking setting. 
Runs were performed on the three datasets, each with one of the three reward signals described in Section~\ref{sec:experiment:setting}, both at the document-level and at the SERP-level. 
Thus there are 27 dataset-model-reward combinations; five runs were performed for each combination from which the mean results are reported. Significance testing was done using a single-tailed T-test.

Parameter tuning was done on the validation sets; the learning rate was tuned between $10^{-2}$ and $10^{-5}$ per dataset; models were updated with early stopping up to 200,000 steps. 
The other parameters were the same for each dataset: the experience replay buffer size was 5,000; the number of steps between transfers was set to 5,000; the exploration degree $\epsilon$ was initialized at $1.0$ and degraded to $0.05$ over 30,000 iterations.
The weights were chosen so that every document representation $\mathbf{\widehat{d}}$ of size $128$, the hidden states $h_t$ of size $256$, and the vectors $v_q$ and $v_p$ have size $128$. 
Gradient descent was performed using Adam optimization with batches of $64$ episodes. All transitions sampled from an episode are included in a batch for computational efficiency.
Our implementation, including parameter settings, are publicly available to ensure reproducibility.

% !TEX root = sigir2018-complex-rankings.tex

\section{Results}
\label{sec:results}

In this section we answer the research questions posed in Section~\ref{sec:intro} by discussing the results from our experiments.
As a reminder, these questions are: do existing methods perform well in complex ranking settings (\ref{rq:existing}); and does \ac{DRM} outperform them in these settings (\ref{rq:drmbetter}). 
Additionally, we also compare the performance of methods under a document-level and SERP-level reward.

Our main results are displayed in Table~\ref{tab:mainresults}, which shows the average \ac{P-NDCG} for each method and reward signal. Figure~\ref{fig:labelpositions} also displays the average relevance label per position for a \ac{GRU} baseline model and \ac{DRM} on the Istella dataset under SERP-level rewards.

\begin{table*}[tb]
\centering
\caption{Average normalized rewards for two baseline models and \ac{DRM} under different complex ranking settings and reward signals. Bold values indicate the best performance per dataset and setting. Significant positive differences are indicated by \dubbelop ($p<0.01$) and \enkelop ($p<0.05$), significant negative differences by \dubbelneer and \enkelneer, respectively. In the table, the first symbol indicates the significance of a difference between \ac{DRM} and MDP-DIV, the second between \ac{DRM} and \ac{GRU}.}
\label{tab:mainresults}
% !TEX root = ../sigir2018-complex-rankings.tex

%\begin{tabular*}{\textwidth}{@{\extracolsep{\fill} } l  l l l l l @{}c@{}c@{}c@{}c }

\newcommand{\dd}{\phantom{\dubbelneer\dubbelneer}}

\setlength{\tabcolsep}{1.95pt}

\begin{tabular}{ l  c c c  c c c  c c c  }
\toprule
 & \multicolumn{3}{c}{\textbf{MSLR}}  & \multicolumn{3}{c}{\textbf{Yahoo}}  & \multicolumn{3}{c}{\textbf{Istella}}  \\
 \cmidrule(r){2-4}
 \cmidrule(r){5-7} 
 \cmidrule(r){8-10} 
  & first-bias & center-bias & last-bias & first-bias & center-bias & last-bias & first-bias & center-bias & last-bias  \\
%\midrule
%& \multicolumn{9}{c}{\textit{Oracle Methods}} \\
%\midrule
%Listwise\\
%Pairwise\\
\midrule
& \multicolumn{9}{c}{\textit{Document-Level Reward}} \\
\midrule
\small
MDP-DIV
& \small 0.423 {\tiny($0.002$) \dd} & \small 0.360 {\tiny($0.004$) \dd} & \small 0.339 {\tiny($0.003$) \dd}
& \small  0.716 {\tiny($0.004$) \dd}  & \small 0.620 {\tiny($0.003$) \dd}  & \small 0.587 {\tiny($0.004$) \dd} 
& \small 0.579 {\tiny($0.003$) \dd}  & \small 0.451 {\tiny($0.002$) \dd} & \small 0.399 {\tiny($0.002$) \dd}\\
\small
GRU
& \small 0.433 {\tiny($0.012$) \dd}  & \small 0.388 {\tiny($0.002$) \dd}  & \small 0.369 {\tiny($0.003$) \dd}
& \small 0.722 {\tiny($0.004$) \dd}  & \small 0.656 {\tiny($0.001$) \dd}  & \small 0.600 {\tiny($0.002$) \dd}
& \small 0.623 {\tiny($0.002$) \dd} & \small 0.532 {\tiny($0.003$) \dd} & \small 0.444 {\tiny($0.002$) \dd} \\
\small
\OurModel 
& \small \bf 0.444 {\tiny($0.004$) \dubbelop$^{-}$} & \small \bf 0.444 {\tiny($0.005$) \dubbelop\dubbelop} & \small \bf 0.445  {\tiny($0.004$) \dubbelop\dubbelop}
& \small \bf 0.728 {\tiny($0.002$) \dubbelop\dubbelop} & \small \bf 0.721 {\tiny($0.001$) \dubbelop\dubbelop} & \small \bf 0.725 {\tiny($0.001$) \dubbelop\dubbelop}
& \small \bf  0.627 {\tiny($0.001$) \dubbelop\enkelop} & \small \bf 0.624 {\tiny($0.003$) \dubbelop\dubbelop} & \small \bf 0.627 {\tiny($0.004$) \dubbelop\dubbelop}\\
\midrule
& \multicolumn{9}{c}{\textit{SERP-Level Reward}} \\
\midrule
\small
MDP-DIV
& \small  \bf 0.414 {\tiny($0.004$) \dd} & \small  0.360 {\tiny($0.006$) \dd} & \small  0.338 {\tiny($0.003$) \dd}
& \small  0.702 {\tiny($0.002$) \dd} & \small  0.611 {\tiny($0.002$) \dd} & \small  0.584 {\tiny($0.000$) \dd}
& \small  \bf 0.549 {\tiny($0.010$) \dd} & \small  0.438 {\tiny($0.005$) \dd} & \small  0.389 {\tiny($0.005$) \dd} \\
\small
GRU
& \small  0.402 {\tiny($0.004$) \dd} & \small  0.367 {\tiny($0.015$) \dd}  & \small  0.344 {\tiny($0.003$) \dd}
& \small  \bf 0.708  {\tiny($0.003$) \dd} & \small  0.625  {\tiny($0.004$) \dd} & \small  0.599  {\tiny($0.003$) \dd}
& \small  0.522 {\tiny($0.007$) \dd}  & 0 \small .449 {\tiny($0.008$) \dd}  & \small  0.422 {\tiny($0.004$) \dd}  \\
\small
\OurModel
& \small  0.395 {\tiny($0.003$) \dubbelneer\dubbelneer} & \small  \bf 0.390 {\tiny($0.016$) \dubbelop\enkelop} & \small  \bf 0.384  {\tiny($0.013$) \dubbelop\dubbelop}
& \small  0.669  {\tiny($0.016$) \dubbelneer\dubbelneer} & \small  \bf 0.635  {\tiny($0.004$) \dubbelop\dubbelop} & \small  \bf 0.647  {\tiny($0.012$) \dubbelop\dubbelop}
& \small  0.533 {\tiny($0.010$) \dubbelop\enkelneer}  & \small  \bf 0.532 {\tiny($0.011$) \dubbelop\dubbelop}  & \small  \bf 0.534 {\tiny($0.007$) \dubbelop\dubbelop}  \\
\bottomrule
%\end{tabular*}
\end{tabular}

\end{table*}

\subsection{Baseline performance} % for Complex Rankings}
First, we answer \ref{rq:existing} by looking at the performance of the MDP-DIV and \ac{GRU} baseline methods. 
Table~\ref{tab:mainresults} shows that across all datasets there is a substantial difference in performance between the ranking settings. 
As expected, both MDP-DIV and \ac{GRU} perform best in the first-bias settings, where their assumed display order matches the preferred order in the setting. 
Conversely, performance is significantly worse in the center-bias and last-bias settings. On the Yahoo and Istella datasets this difference can go up to $0.1$ \ac{P-NDCG}, under both document-level and SERP-level rewards.
This shows that the performance of both baselines is heavily affected by the preferred display order, and in situations where this order is unknown this can lead to a substantial drop in performance.
 
To better understand the drop in performance of the \ac{GRU} baseline, we look at Figure~\ref{fig:grudistribution}.
Here we see that the reward signals do affect the ranking behaviour the \ac{GRU} learns.
For instance, in the center-bias case the most preferred display positions ($p_5$) has the highest average relevance label.
However, the complete preferred display order does not seem to be inferred correctly.
Notably, the relevance on the first three positions ($p_1, p_2, p_3$) is compromised, i.e., position $p_3$ is preferred over $p_8$ but has a much lower average relevance label.
It seems that \ac{GRU} intentionally ranks worse at the start of the ranking so that the most relevant documents are not placed too early.
This behavior places the most relevant document on the most preferred position, but sacrifices the relevance at earlier positions.
Thus, the \ac{GRU} baseline is unable to rank for the center-bias display order.
Moreover, in the last-bias setting $p_3$ has the highest average relevance while $p_{10}$ is the most preferred display position.
\ac{GRU} appears to be unable to follow this preferred display order and, instead, optimizes for the relevance order only.

Due to the significant decreases in performance that are observed for the MDP-DIV and \ac{GRU} baseline models in different ranking settings, we answer \ref{rq:existing} negatively and conclude that neither baseline model performs well in complex ranking settings.

\begin{figure*}[tb]
\centering
\addtocounter{figure}{1}
\begin{subfigure}{\textwidth}
%\vspace{-.5\baselineskip}
\includegraphics[scale=0.27]{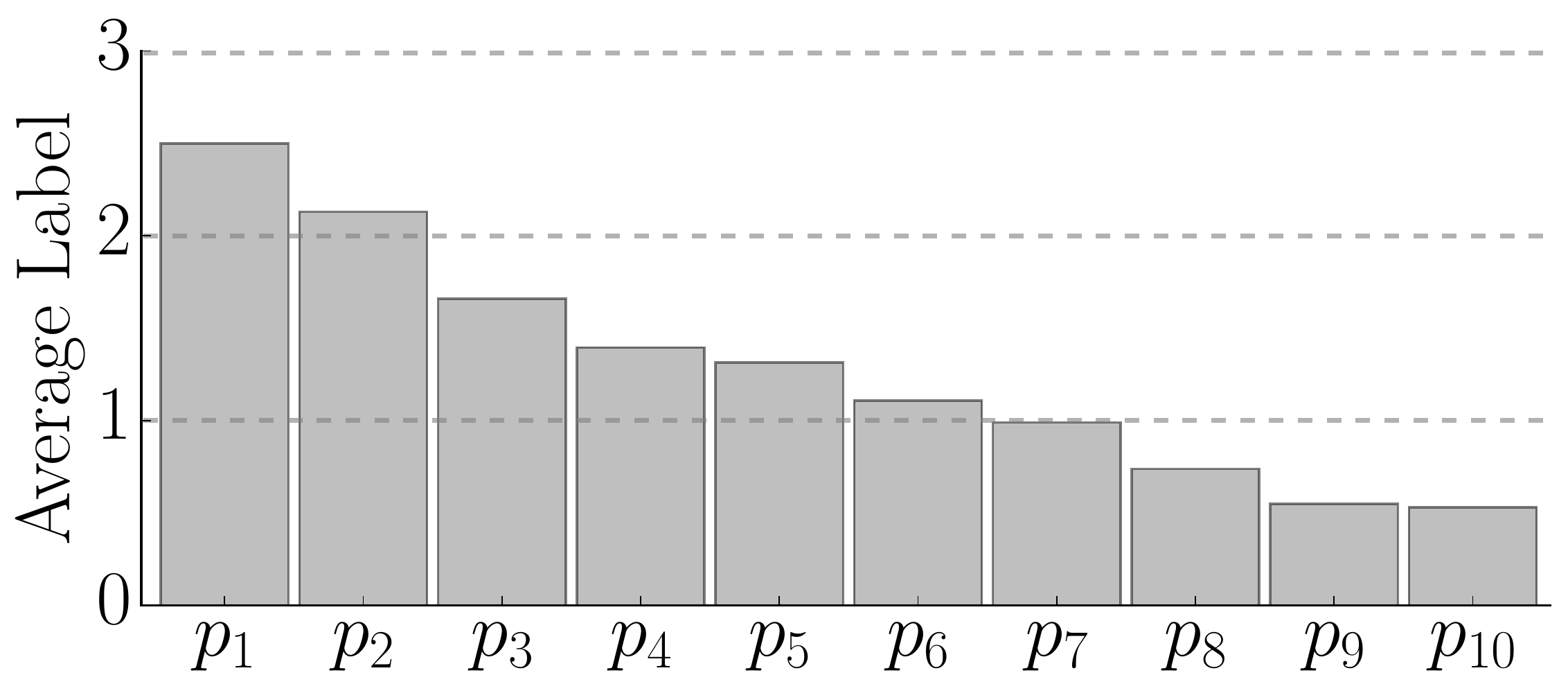}
\includegraphics[scale=0.27]{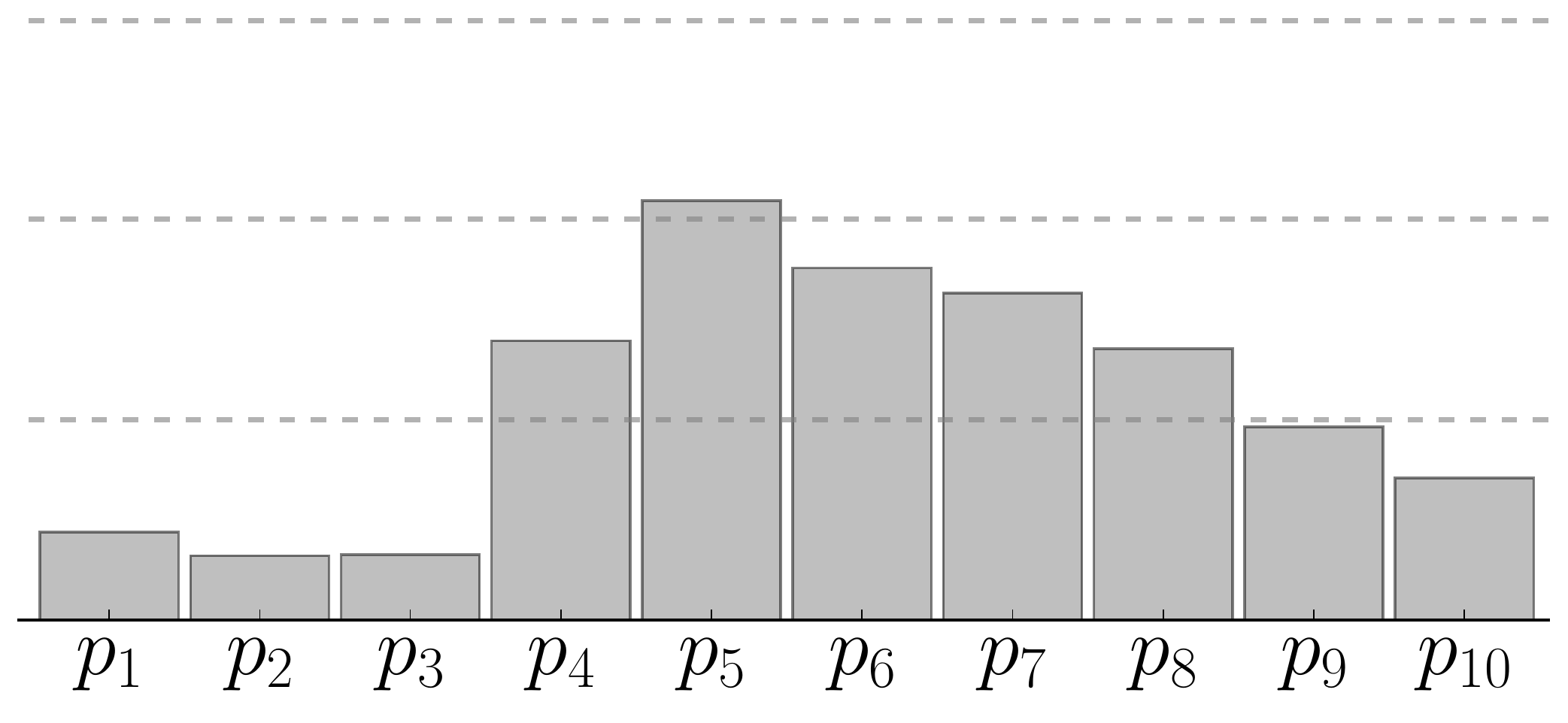}
\includegraphics[scale=0.27]{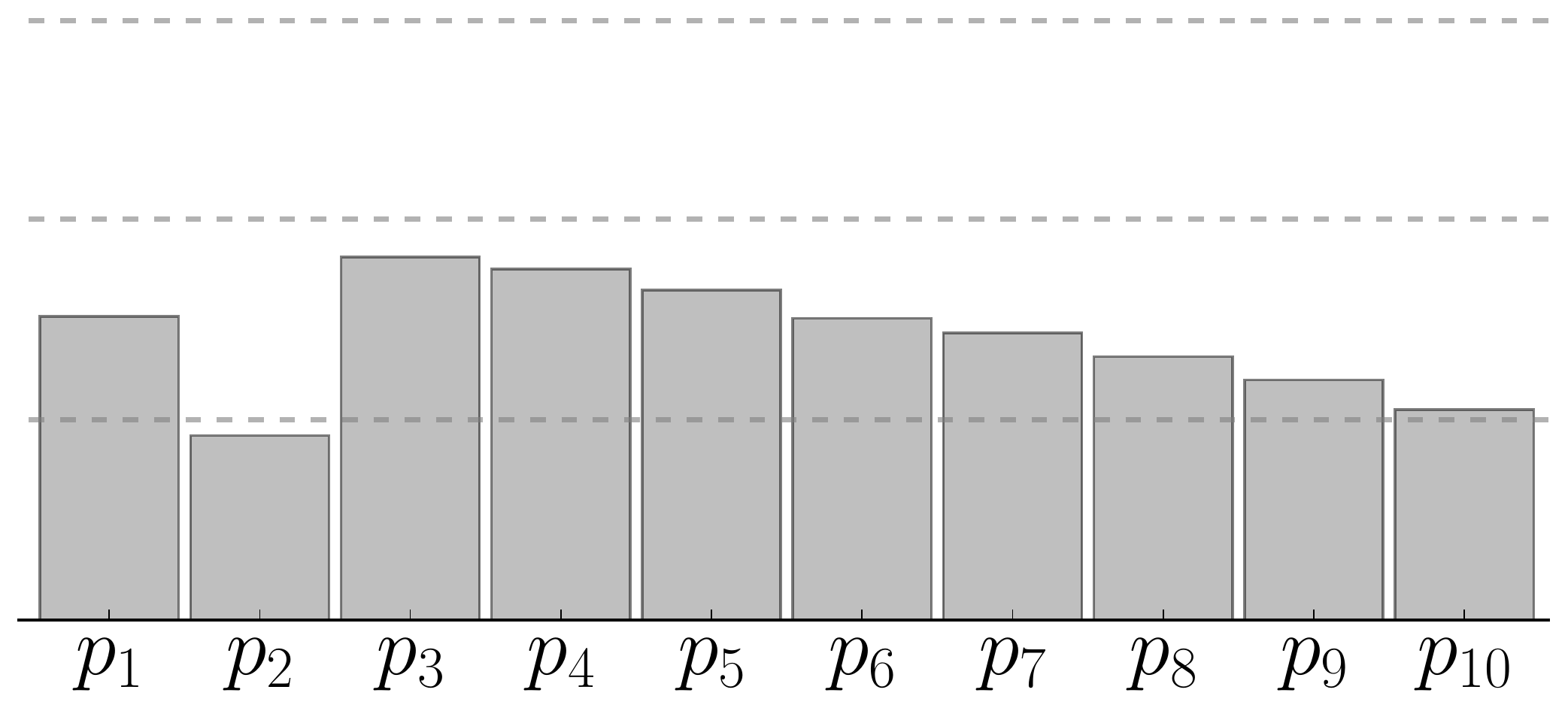}
\caption{Average label per position for the \ac{GRU} baseline model in different settings: first-bias(left), center-bias (middle) and last-bias (right).}
\label{fig:grudistribution}
\end{subfigure}
%\vspace{-2.5\baselineskip}
\begin{subfigure}{\textwidth}
%\vspace{-.5\baselineskip}
\includegraphics[scale=0.27]{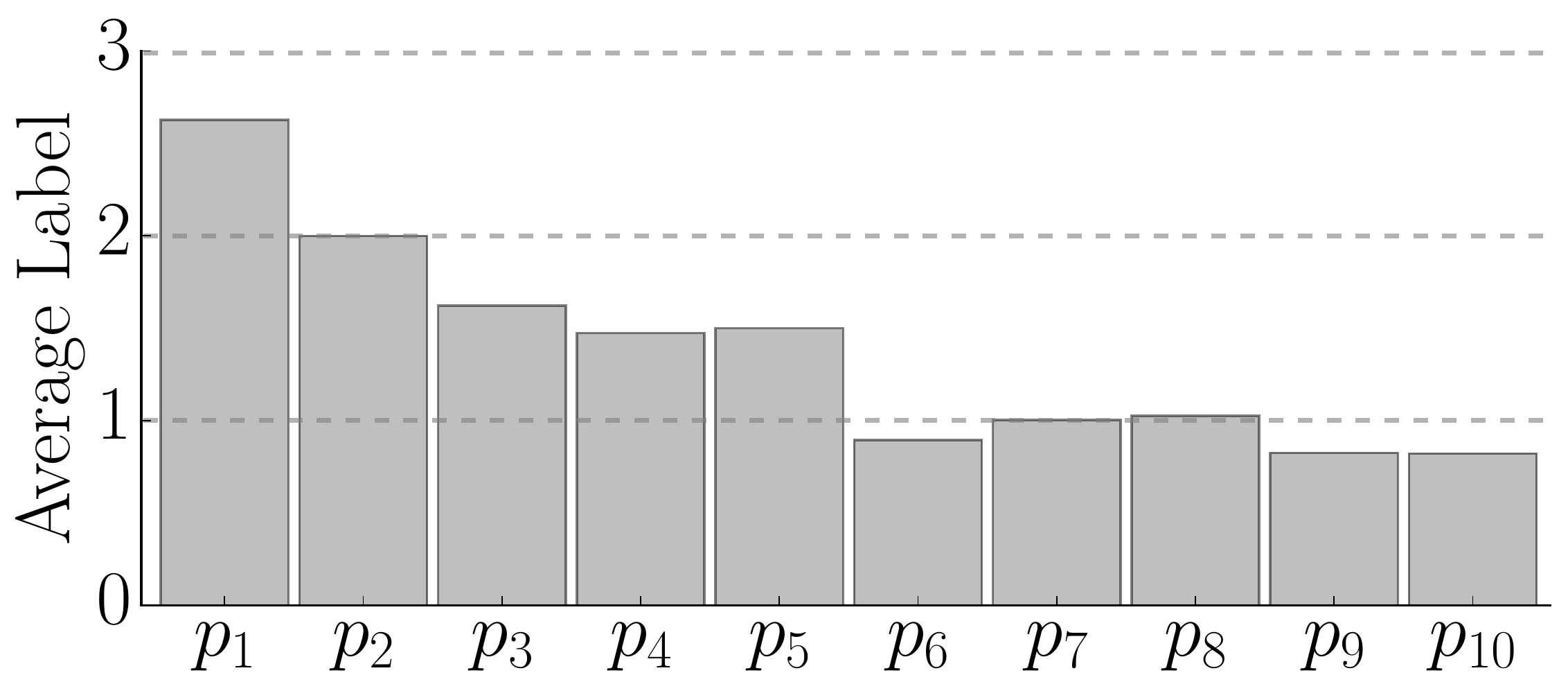}
\includegraphics[scale=0.27]{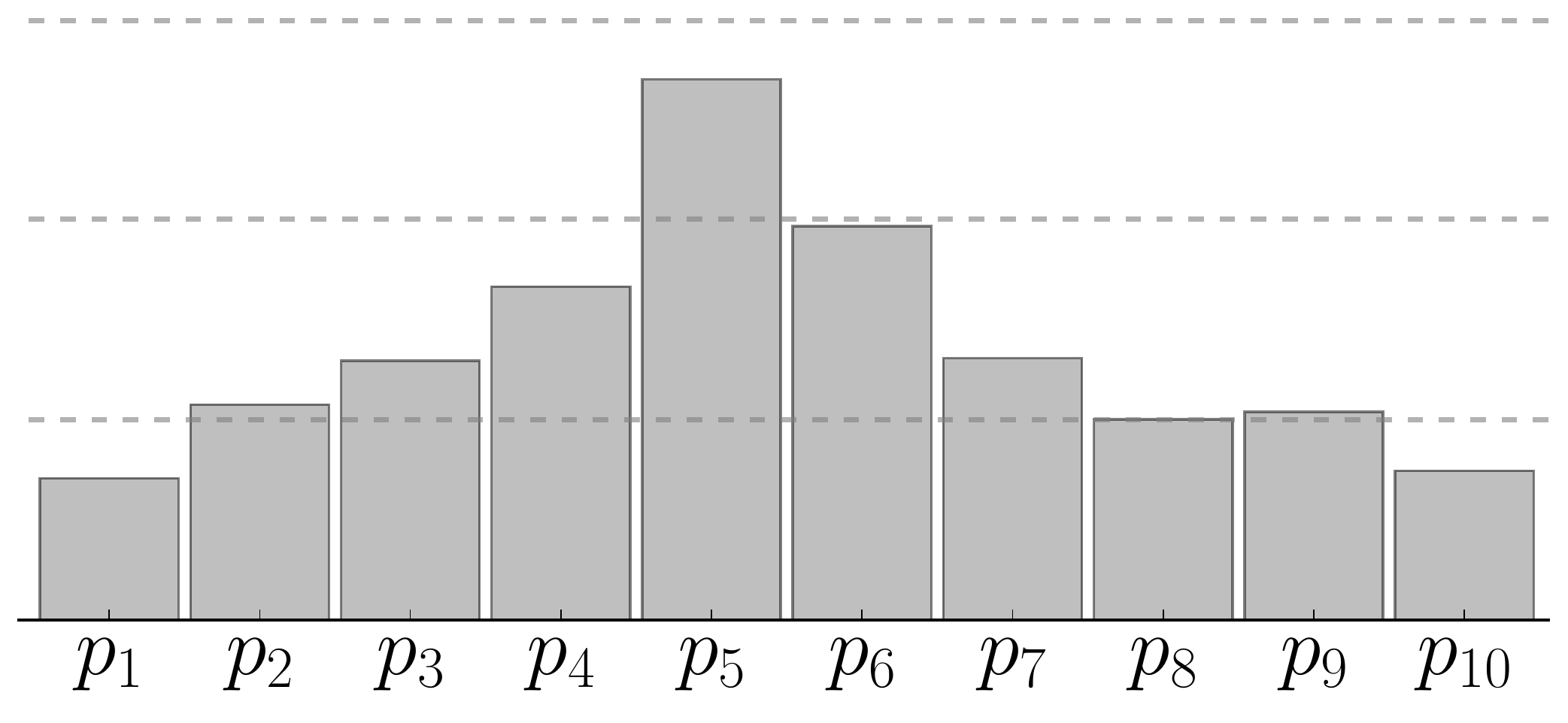}
\includegraphics[scale=0.27]{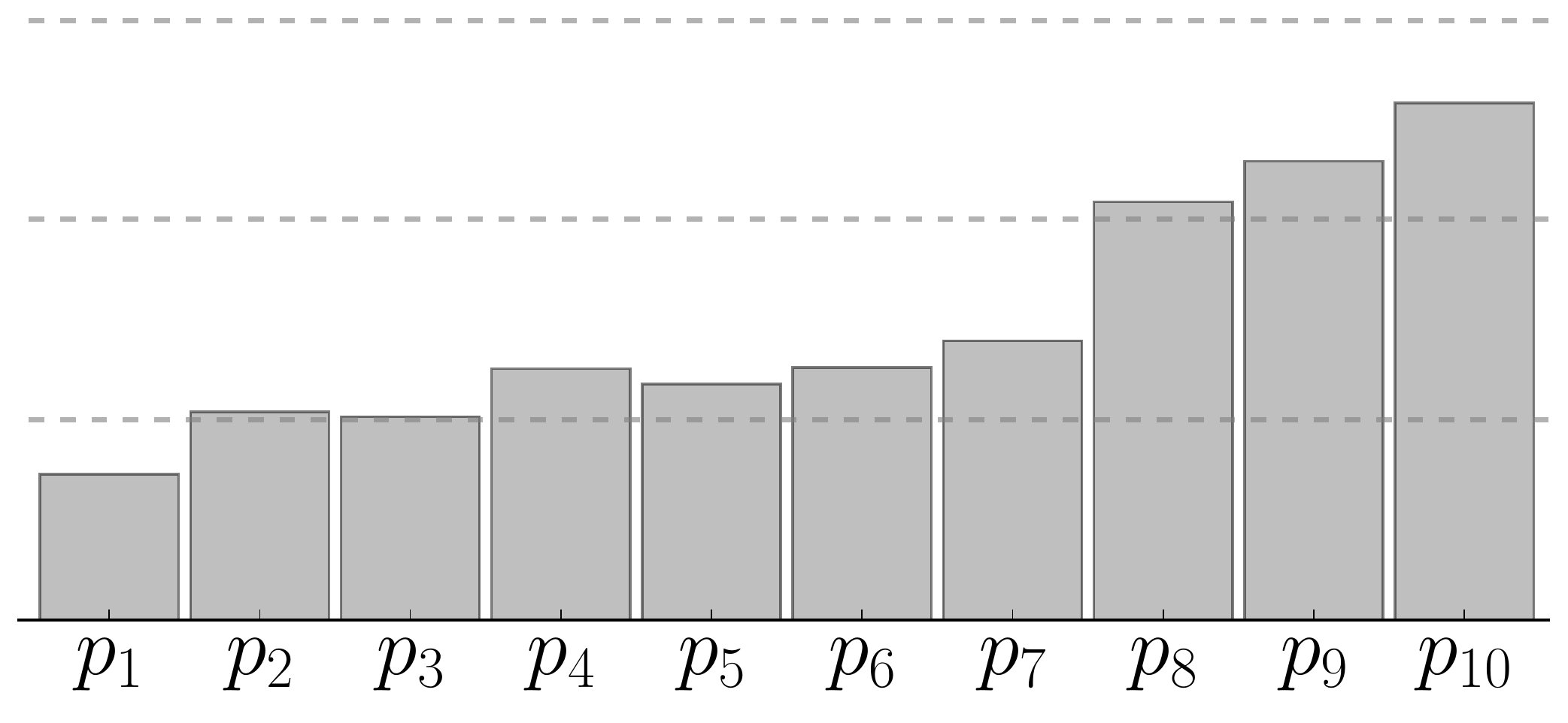}
\caption{Average label per position for the \ac{DRM} in different settings: first-bias(left), center-bias (middle) and last-bias (right).}
\label{fig:drmdistribution}
\end{subfigure}
%\vspace{-2.5\baselineskip}
\begin{subfigure}{\textwidth}
%\vspace{-.5\baselineskip}
\includegraphics[scale=0.27]{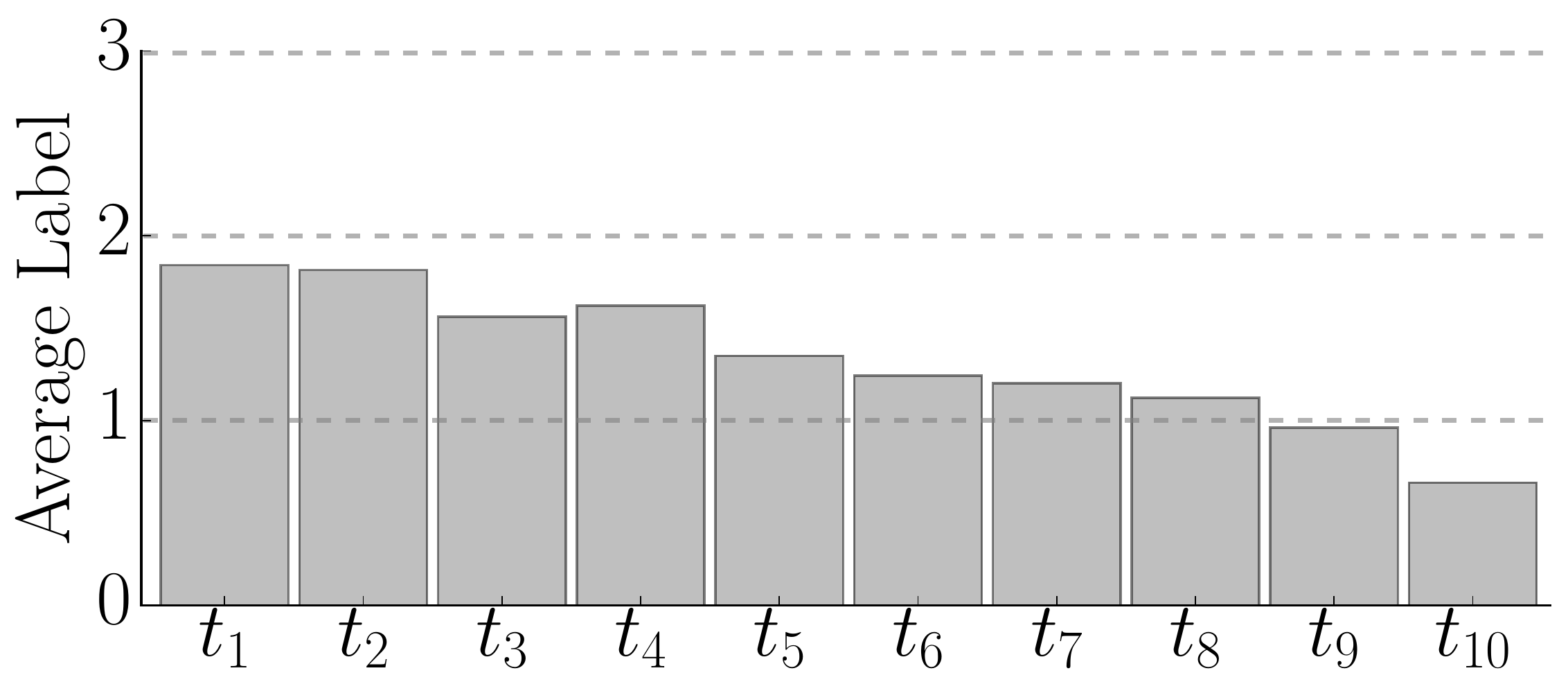}
\includegraphics[scale=0.27]{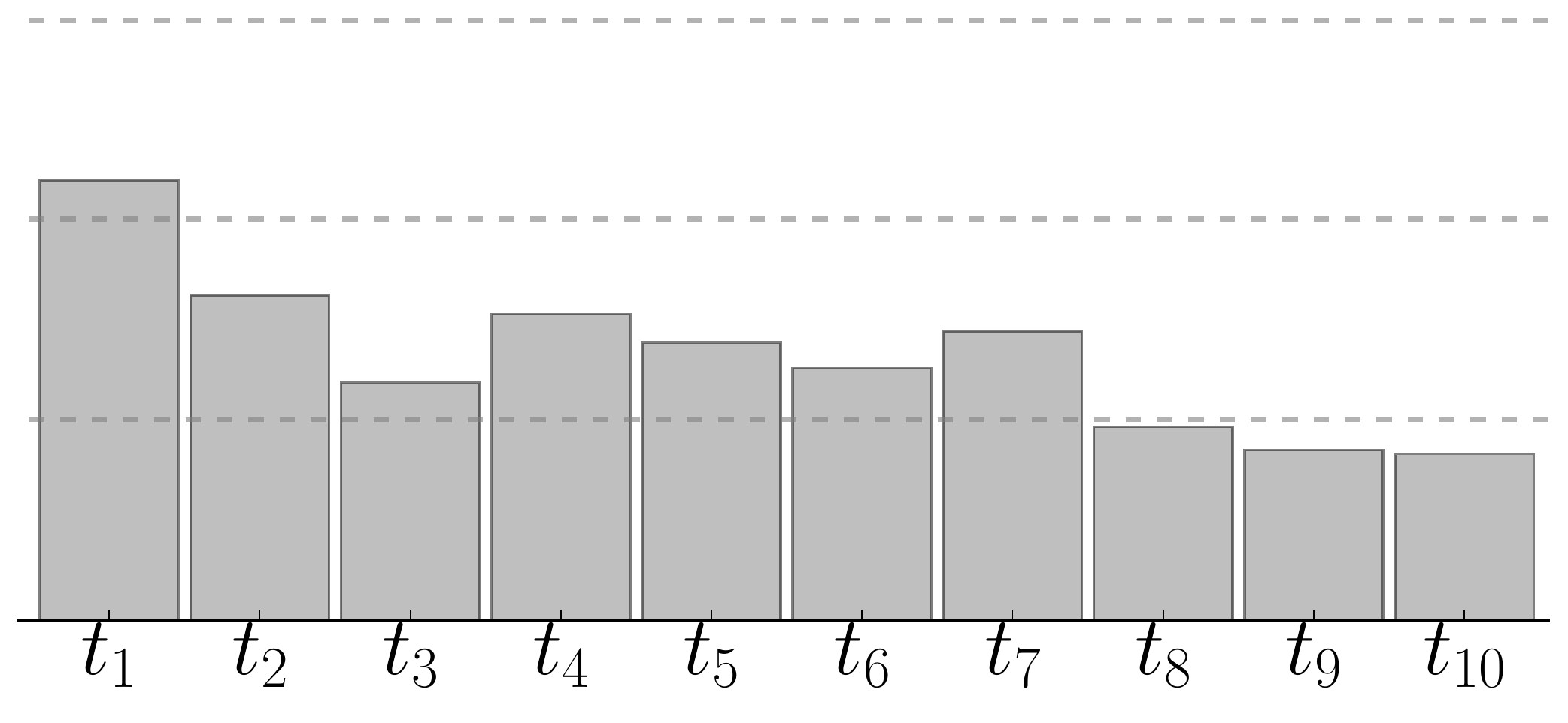}
\includegraphics[scale=0.27]{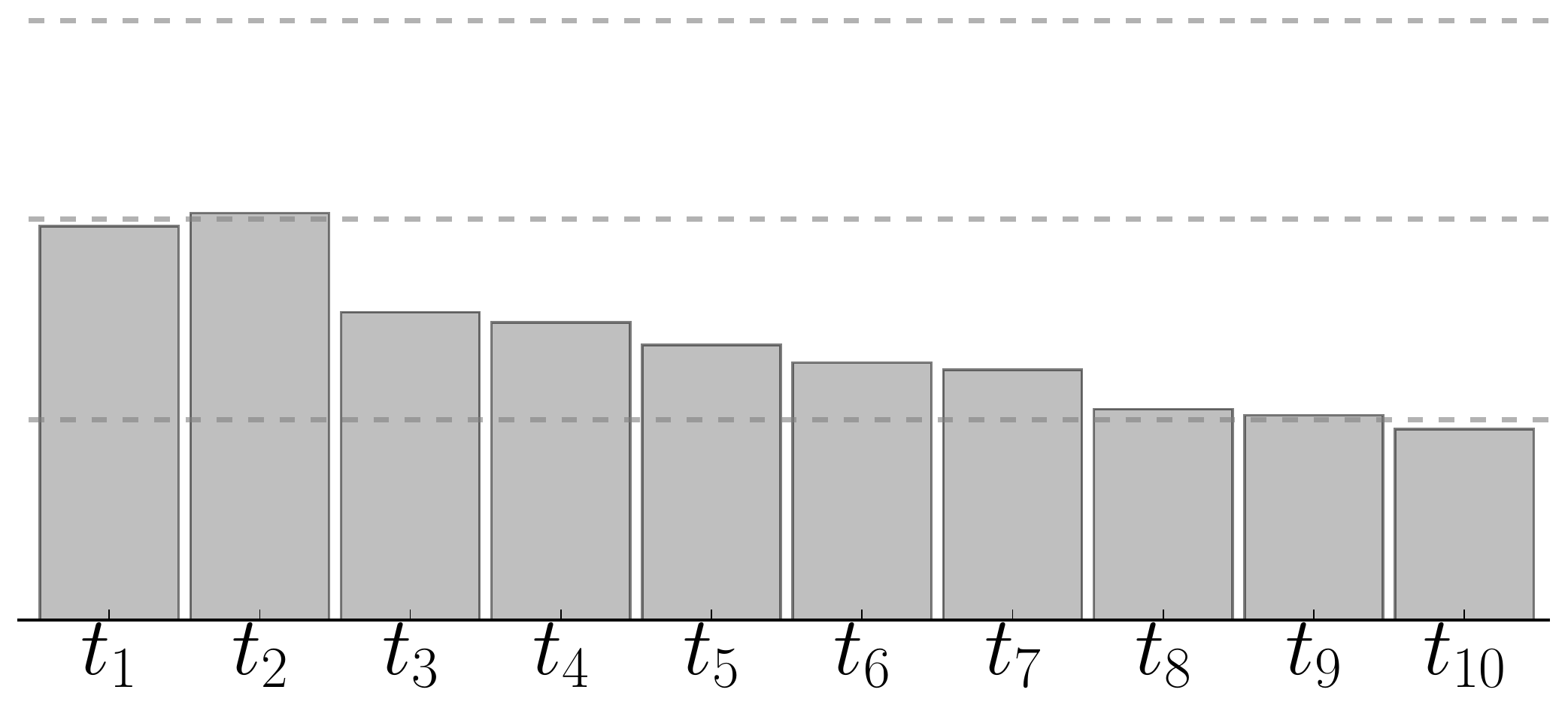}
%%% and correct
\caption{Average label per time-step for the \ac{DRM} in different settings: first-bias(left), center-bias (middle) and last-bias (right).}
\label{fig:drmtimestep}
\end{subfigure}
%\vspace*{-\baselineskip}
\caption{Average label per position for a single \ac{GRU} baseline model and a single \ac{DRM} trained on the Istella dataset in different complex ranking settings under SERP-level rewards.}
\label{fig:labelpositions}
%\vspace*{-\baselineskip}
\end{figure*}

\subsection{\ac{DRM} performance on complex rankings}
In order to answer \ref{rq:drmbetter}, we consider the performance of \ac{DRM} reported in Table~\ref{tab:mainresults}.
In the first-bias setting the \ac{DRM} performs similarly to the baseline models across all datasets.
While under SERP-level rewards the performance varies per dataset, under document-level rewards \ac{DRM} performs better than the baseline methods in all cases.
The performance of \ac{DRM} in the center-bias and last-bias settings is significantly better than the baseline models, with a substantial difference of over $0.1$ \ac{P-NDCG} in multiple cases.
Thus, if there is a SERP-level reward signal and the user preferred display order is non-standard then \ac{DRM} significantly outperforms the baseline methods.
Under decomposable rewards \ac{DRM} achieves the best ranking performance regardless of the preferred display order.

Furthermore, Table~\ref{tab:mainresults} shows that the performance of \ac{DRM} is not affected by the preferred display order.
Unlike the baseline methods, \ac{DRM} performs similarly in the first-bias, center-bias and last-bias settings.
\ac{DRM} differs from the baselines by not having an assumed display order, and therefore it also does not have a disadvantage in any of the settings.
Thus, by being able to determine the order in which positions are filled, \ac{DRM} can maintain its performance in any complex ranking setting.
This is further illustrated in Figure~\ref{fig:drmdistribution}: \ac{DRM} correctly identifies the preferred display order for each complex ranking setting. 
In all three cases, the more preferred positions have a higher average relevance label, with only a few minor errors in the inferred order of less preferred positions.
Then, Figure~\ref{fig:drmtimestep} displays the average relevance label of the document \ac{DRM} selects per time-step; in all settings, the first placed documents are the most relevant.
\ac{DRM} follows the preferred relevance order when selecting documents, but places them in positions according to the preferred display order.
Thus, as hypothesized, \ac{DRM} does not \emph{save} relevant documents for later time-steps; we attribute the better performance of \ac{DRM} to this difference.

In conclusion, we answer \ref{rq:drmbetter} positively. 
\ac{DRM} outperforms the baseline methods in settings where the preferred display order is unknown. 
By looking at the average relevance label per position, we see that \ac{DRM} correctly infers the preferred order in all cases. 
Consequently, by correctly inferring both the relevance and display order \ac{DRM} performs consistently well in all complex ranking settings.

\subsection{Document-level vs.\ SERP-level rewards}
Lastly, we briefly discuss the difference in performance between document-level rewards and SERP-level rewards.
Table~\ref{tab:mainresults} shows that all methods perform better under document-level rewards.
This is expected since under SERP-level rewards the methods have to infer how the reward signal is decomposed, while with document-level rewards this decomposition is directly observed.
Under SERP-level rewards in the first-bias setting, it is also less clear what the best performing method is, as it varies per dataset.
Conversely, under document-level rewards the \ac{DRM} method reaches the highest performance across all datasets.
However, despite all methods performing less well under SERP-level rewards, the \ac{DRM} still has the best performance on the center-bias and last-bias settings.
Figure~\ref{fig:drmdistribution} shows that the \ac{DRM} is capable of inferring the correct display order from the less informative SERP-level reward signal.
Thus, even with very weak reward signals \ac{DRM} is robust to display preferences in complex ranking settings.

%\todo{Moeten we geen significantie tests opnemen bij ``bake-offs'' tussen methodes/reward signals?}

%Lastly, we consider the difference between document-level rewards and SERP-level rewards to answer \ref{rq:serpreward}. Table~\ref{tab:mainresults} shows that all methods perform better under document-level rewards. This is expected since under the SERP-level rewards the methods have to infer how the reward signal is decomposed. With the document-level rewards this decomposition is directly observed. However, under SERP-level rewards the same trends are seen: \ac{DRM} is not affected by different ranking settings while the baselines are. We conclude that \ac{DRM} is still effective when the reward signal is not decomposable.

% !TEX root = sigir2018-complex-rankings.tex

\section{Related Work}
\label{sec:related}

We discuss research related to complex ranking settings: whole page optimization and ranking for alternative presentations formats. 

\subsection{Whole page optimization}
In recent years, the \emph{ten blue links} format has become less prominent and search engines now usually display a selection of images, snippets and information boxes along their search results \citep{wang2016beyond}.
As SERPs contain more alternative items, their evaluation has changed accordingly.
Previous work has looked at whole page evaluation, where either human judges \citep{bailey2010evaluating} or user models \citep{chuklin-incorporating-2016} are used to evaluate the entire SERP.
Such user models consider the possibility that users may direct their attention to information boxes or snippets in the SERP.
As a result, they can recognize that some forms of user abandonment can be positive, because the user can be satisfied before clicking any links.

\citet{wang2016beyond} introduced the idea of \emph{whole page presentation optimization}, where the entire result presentation is optimized, including item position, image size, text font and other style choices.
While this method addresses the importance of user presentation preferences, it does not decide what items to display, but assumes this set is given.
%Thus it does not decide what items should be displayed only how they are displayed.
The method assumes that the number of possible presentation formats is small enough to iterate over.
In the complex ranking setting this would lead to combinatorial problems as for a layout with $k$ positions there are $k!$ preferred display orders possible.
While this work optimizes result presentation, it does not consider the mismatch between the user preferred relevance order and display order.

\subsection{Alternative presentation formats}

Section~\ref{sec:intro} has already discussed a number of different search settings and layouts. 
While this paper has focussed on optimizing for the relevance and display preferences of users, previous work has looked at optimizing ranking for specific alternative presentations.
For instance, SERPs in mobile search generally use single column vertical layouts for the relatively small mobile screens.
\citet{luo2017evaluating} recognize that the percentage of the screen an item covers should be taken into evaluation and propose an evaluation method that discounts items according to the varying lengths.

In addition, existing work on optimizing display advertisement placement has noted that presentation context can have a big impact on the click-through-rate of an ad. 
For instance, \citet{metrikov2014whole} found that whole-page-satisfaction leads to higher click-through-rates on the ad block of SERPs. 
Furthermore, ads can be more effective if they are optimized to display multiple products at once \citep{lefortier-large-scale-2016}, or if multiple banners display the related advertisements \citep{devanur2016whole-page}.

Similar to the concept of complex ranking settings, these existing studies show that for the best user experience the presentation of a ranking should be optimized together with its content.
% !TEX root = sigir2018-complex-rankings.tex

\section{Conclusion}
\label{sec:conclusion}

In this paper we have formalized the concept of a complex ranking setting, where the user has a preferred relevance order -- an ordering of documents -- and a preferred display order -- a preference concerning the positions in which the documents are displayed.
The ideal ranking in a complex ranking setting matches the relevance order with the display order, i.e., where the most relevant document is displayed in the most preferred position, and so forth.
Thus these settings pose a dual problem: for the optimal user experience both the relevance order and the display order must be inferred.
We simulate this setting by performing experiments where a permuted DCG is given as a reward signal, e.g., in the last-bias ranking setting the last display position is discounted as if it was the most preferred position according to the user.

Our experiments show that existing methods are unable to handle the complex ranking setting. 
Our results show that while existing methods learn different behaviors in different settings, they are unable to rank documents according to any non-standard preferred display order.
As an alternative, we have introduced \ac{DRM}, which sequentially selects documents and the positions to place them on.
By explicitly modeling the duality between the relevance order and display order, \ac{DRM} is able to infer the user preferred display order and relevance order simultaneously.
Our results show that \ac{DRM} matches state-of-the-art performance in settings where the display order is known, and significantly outperforms previous methods in complex ranking settings.
Thus, when dealing with complex presentation layouts where it is unclear in what order users consider display positions, \ac{DRM} is able to provide good performance regardless of the users' actual display preferences.

Future work could consider more noisy reward signals; while our setup simulates complex ranking settings, it does so without noise in the reward signal.
Even under these ideal circumstances existing methods are unable to infer non-standard preferred display orders.
It would be interesting to investigate whether \ac{DRM} is still able to handle complex ranking settings with high levels of noise.
Lastly, the deployment of \ac{DRM} on an actual search setting could show the impact display preferences have on user experiences.

\subsection*{Code}
To facilitate reproducibility of the results in this paper, we are sharing the code used to run the experiments in this paper at \url{https://github.com/HarrieO/RankingComplexLayouts}.

\subsection*{Acknowledgements}
This research was supported by
Ahold Delhaize,
Amsterdam Data Science,
the Bloomberg Research Grant program,
the China Scholarship Council,
the Criteo Faculty Research Award program,
Elsevier,
the European Community's Seventh Framework Programme (FP7/2007-2013) under
grant agreement nr 312827 (VOX-Pol),
the Google Faculty Research Awards program,
the Microsoft Research Ph.D.\ program,
the Netherlands Institute for Sound and Vision,
the Netherlands Organisation for Scientific Research (NWO)
under pro\-ject nrs
CI-14-25, % MediaNow
652.\-002.\-001, % Re-Search
612.\-001.\-551, % CLEAR
652.\-001.\-003, % NEWEL
and
Yandex.
All content represents the opinion of the authors, which is not necessarily shared or endorsed by their respective employers and/or sponsors.

\balance

\bibliographystyle{ACM-Reference-Format}
\bibliography{sigir2018-complex-rankings.bib}

\end{document}